\documentclass{article}

\usepackage{PRIMEarxiv}

\usepackage[utf8]{inputenc} 
\usepackage[T1]{fontenc}    
\usepackage{hyperref}       
\usepackage{url}            
\usepackage{booktabs}       
\usepackage{amsfonts}       
\usepackage{nicefrac}       
\usepackage{microtype}      
\usepackage{lipsum}
\usepackage{fancyhdr}       
\usepackage{graphicx}       
\graphicspath{{media/}}     

\usepackage{algorithm}
\usepackage{algorithmic}
\usepackage{amsmath}
\usepackage{amssymb}
\usepackage{multirow}
\usepackage[table]{xcolor}
\usepackage{subcaption}
\allowdisplaybreaks 

\title{FedLSG: LLM-Enhanced Semantic Calibration for \\ Federated Graph Backdoor Defense
}

\author{
  Chenyu Zhou\\
  Southeast University \\
  Purple Mountain Laboratories \\
  \texttt{zhouchenyu@seu.edu.cn} \\
\And
  Yabin Peng\\
  Southeast University \\
  Purple Mountain Laboratories \\
  \texttt{yabinpeng@seu.edu.cn} \\
\And
  Wei Huang\\
  Institute of AI for Industries \\
  Chinese Academy of Sciences \\
  \texttt{whuang@iaii.ac.cn} \\
\And
  Kunlin Li\\
  Southeast University \\
  Purple Mountain Laboratories \\
  \texttt{230259669@seu.edu.cn} \\
\And
  Shuaishuai Zhang\\
  Purple Mountain Laboratories \\
  \texttt{zhangshuaishuai@pmlabs.com.cn} \\
\And
  Xinyuan Miao \\
  Purple Mountain Laboratories \\
  \texttt{miaoxinyuan@pmlabs.com.cn} \\
}

\begin{document}
\maketitle

\begin{abstract}
Federated Graph Neural Networks (FedGNNs) are highly vulnerable to backdoor poisoning, yet existing defenses typically rely on rule-based approaches that lack semantic understanding, making them vulnerable to stealthy triggers and harmful to benign structures. To solve this, we present FedLSG, the first framework that integrates large language models (LLMs) into federated graph backdoor defense. FedLSG introduces a graph and behavior to text grounding scheme that transforms local graph structures and client update behaviors into semantically rich natural language representations. The framework further adopts a lightweight student–teacher architecture. On the server side, a full scale LLM serves as a teacher, providing global contextual guidance and evaluating client updates during aggregation to identify potentially malicious participants. On the client side, a LoRA-based student is maintained to perform semantic reasoning, to suppress the influence of edges associated with backdoor triggers. By enabling semantic interpretation of both graph  patterns and client behaviors, the framework adaptively incorporates rule-based signals into message passing and client aggregation for defense. Experiments demonstrate that FedLSG significantly improves resistance to backdoor attacks without compromising graph integrity.
\end{abstract}


\section{Introduction}
Graph Neural Networks (GNNs), designed for learning from graph-structured data, have achieved widespread success in applications ranging from recommender systems \cite{lv25dy} to social networks \cite{Li1056} and node classification \cite{zhu25re}. However, centralized GNN training raises significant privacy concerns, motivating the development of Federated Graph Neural Networks (FedGNNs) that enable privacy-preserving distributed training. Despite their privacy advantages, recent studies \cite{Bkdfedgnn,Mostly} reveal that FedGNNs are highly vulnerable to data-poisoning-based backdoor attacks. Attacks such as SBA \cite{SBA}, UGBA \cite{UGBA}, and DPGBA \cite{DPGBA} inject imperceptible trigger subgraphs into local training data of compromised clients. Through federated aggregation, these poisoned local models contaminate the global model, which subsequently propagates backdoor information to all participating clients in future rounds, creating a cascading failure that compromises the entire federation \cite{CAI24FL,NGU24Ba,Pmlr22zh}.

Defending against this threat is substantially harder than defending centralized GNNs. Existing rule-based defenses largely rely on pruning suspicious edges or perturbing graph structure \cite{UGBA,NoisyGCN,RIGBD}. While effective in some centralized settings, they are difficult to deploy in FedGNNs because the server cannot identify which clients are compromised. Applying aggressive filtering to every client often removes useful benign structure and harms clean accuracy, whereas conservative filtering frequently fails to eliminate stealthy triggers \cite{Shen24Be}. Aggregation-based defenses \cite{RLR,FedTGE} attempt to suppress malicious clients during model fusion by setting some rules, but they do not fully solve the problem either: poisoned information that survives aggregation can still be absorbed by benign clients from the global model and accumulate over rounds \cite{CAI24FL,NGU24Ba,Pmlr22zh}. Therefore, robust defense in FedGNNs requires a joint treatment of local suspicious structures, client-global interaction, and server-side aggregation.

These limitations expose three tightly coupled challenges. First, trigger-related edges are highly similar to normal graph connections, rendering anomaly-based pruning unreliable, as it may either remove benign structures and harm clean accuracy or fail to eliminate stealthy triggers. Second, rule structural judgments of edge reliability are inherently limited, as they rely solely on topology signals without incorporating semantic context from the graph, which can lead to unreliable assessment of edge authenticity. Third, once misclassified updates participate in federated aggregation, the resulting poisoned knowledge can be incorporated into the global model and repeatedly propagated to benign clients. Therefore, a robust defense should not rely solely on rule-based structural signals, but should integrate them into semantic-aware signals that support reliable discrimination between benign patterns and backdoor behaviors.

This observation highlights the potential of incorporating LLMs into FedGNNs, as their semantic reasoning capability can complement structural learning and provide global context for edge reliability and backdoor detection. Crucially, LLM-based semantic signals can also guide the aggregation process by offering semantic consistency cues for evaluating client updates, thereby helping to mitigate the propagation of poisoned information across the federation. However, several challenges remain. First, it is non-trivial to transform graph structures and client behaviors into textual representations that are suitable for LLM reasoning. Second, relying solely on either semantic reasoning or rule-based judgments is insufficient in practice, and a principled integration of both is required. Third, deploying full-scale LLMs on every client is often computationally prohibitive in realistic federated settings.

Based on this insight, this paper proposes FedLSG, which integrates LLM-guided semantic reasoning into FedGNNs. Specifically, FedLSG converts graph structures, including neighborhood dependencies, anomalous structural patterns, and connectivity irregularities, into textual descriptions, enabling the LLM to assess whether edges are potentially abnormal. Furthermore, FedLSG transforms client training dynamics, such as model update magnitudes, gradient variation trends, and deviations from the global model, into textual descriptions, allowing the LLM to reason about whether a client behaves anomalously.

Moreover, to reduce complexity, FedLSG introduces a lightweight student–teacher architecture. It builds on a GAT-based structural backbone, which performs adaptive edge-weighted message passing under the guidance of LLM, to distinguish normal connections from trigger-related ones. The server maintains a global GAT and a full LLM as the teacher, while each client employs a GAT with a lightweight LoRA-based student to learn knowledge distilled from the server-side LLM.

At the server side, a semantic teacher leverages full-scale LLM reasoning to provide global contextual guidance and shared semantic priors across clients. In addition, the LLM is further used to analyze client model updates at the behavioral level, such as whether updates are consistent with global semantic patterns or exhibit abnormal clustering or deviation trends. By adaptively combining semantic reasoning with rule-based judgments, the server performs trust evaluation during aggregation, allowing it to detect malicious clients and suppress their impact to the global model.

At the client side, a lightweight LoRA-based student is used to model local graph semantics by distilling global semantic priors from the LLM teacher. It assesses whether edges are semantically aligned with the local structural context, e.g., whether connected nodes are consistent in semantic roles, functional properties, or structural contexts. The resulting semantic signal is combined with rule-based signals, and injected into GAT to dynamically adjust edge weights. This allows potential backdoor edges to be down-weighted during message passing instead of being removed, ensuring robust propagation while preserving graph structure.


We conduct extensive experiments on four benchmark datasets under three state-of-the-art attacks. Results demonstrate that FedLSG achieves the lowest attack success rate across all configurations while maintaining competitive clean accuracy. We further validate FedLSG's robustness under adaptive white-box attacks, varying numbers of malicious clients, LLM backbone selection, large-scale client settings, heterogeneous graphs, and IID data partitions.

\section{Related Work}

\paragraph{Backdoor Attacks \& Defenses in FedGNNs.}  Recent studies have shown that GNNs are highly vulnerable to backdoor attacks. SBA \cite{SBA} injects subgraph triggers into training graphs without requiring knowledge of model structure. UGBA \cite{UGBA} generates adaptive triggers that closely resemble target nodes under a limited budget. DPGBA \cite{DPGBA} further improves stealthiness via intra-distributed trigger design. Bkd-FedGNN \cite{Bkdfedgnn} presents a backdoor attack benchmark for FedGNNs, integrating above backdoor attacks into a unified framework. In response, several defense methods have been proposed. Prune \cite{UGBA} disables subgraph triggers by pruning edges between low-similarity nodes. RIGBD \cite{RIGBD} introduces randomized edge-dropping and robust training to reduce trigger damage. FedTGE \cite{FedTGE} employs an energy-aware mechanism to identify clean and attacked samples, and accordingly distinguishes clients during the aggregation phase.

\paragraph{LLM in FedGNNs.} Recent studies explore integrating LLM into FedGNNs to enhance structural and semantic reasoning.  LLM4FGL \cite{LLM4FGL} generates node textual descriptions and infers missing edges between nodes to mitigate data heterogeneity. LG-DUMAP \cite{LGDUMAP} further incorporates instruction-tuned encoders and in-context examples for node classification and link prediction. However, these methods require deploying full LLMs on all clients, incurring high computational and communication costs. To address this issue, pFedLoRA \cite{pFedLoRA} combines FL with LoRA-based efficient tuning for lightweight client-side adaptation. And FedAMoLE \cite{FedAMoLE} personalizes client based on local data, FedEx-LoRA \cite{FedExLoRA} improves aggregation with residual errors, and FedALT \cite{FedALT} integrates shared knowledge via independent LoRA without aggregated initialization.


\section{Preliminaries}

\paragraph{Training of FedGNN.} Consider a FedGNN system comprising $N_c$ clients and a central server, where each participant maintains a local GNN model. Privacy constraints necessitate partitioning the original graph $G$ into multiple subgraphs $\{G_i\}_{i=1}^{N_c}$ distributed across clients, precluding centralized training \cite{Wa22Fg}. The FedGNN training protocol proceeds iteratively. At iteration $t$, the following steps are executed \cite{Bkdfedgnn}:

\textbf{(1) Model Distribution:} The server broadcasts the global model parameters $\theta^t$ to all participating clients.

\textbf{(2) Local Update:} Each client $i$ performs local optimization on its private subgraph $G_i$ by minimizing the loss function $L$ with learning rate $\eta$:
\begin{equation} \label{Eq_loc}
\theta_i^{t} = \theta^t - \eta \nabla_{\theta} L(\mathcal{G}_i; \theta^t).
\end{equation}

\textbf{(3) Model Aggregation:} The server aggregates the received local models using a predefined strategy to update the global model:
\begin{equation} \label{Eq_glo}
\theta^{t+1} = \frac{1}{N_c}\sum_{i=1}^{N_c} \theta_i^{t}.
\end{equation}

\paragraph{Backdoor Attacks in FedGNNs.} In backdoor attacks against federated graph learning, an adversary compromises client $i$ by injecting a trigger subgraph $G'_i$ into the local graph $G_i$ \cite{Bkdfedgnn}. Training on this poisoned data produces a backdoored local model $\tilde{\theta}_i^{t}$ that contaminates the global model $\theta^{t+1}$ during aggregation, subsequently propagating the backdoor to benign clients in future rounds. The threat model assumes the adversary can only manipulate local data of compromised clients without interfering with other clients' training or the server's aggregation process. Additionally, the adversary may dynamically select which clients to compromise across different rounds, and the server cannot distinguish between benign and compromised clients, making the attack both flexible and stealthy.

\begin{figure}[]
\centering
\includegraphics[width=0.95\columnwidth]{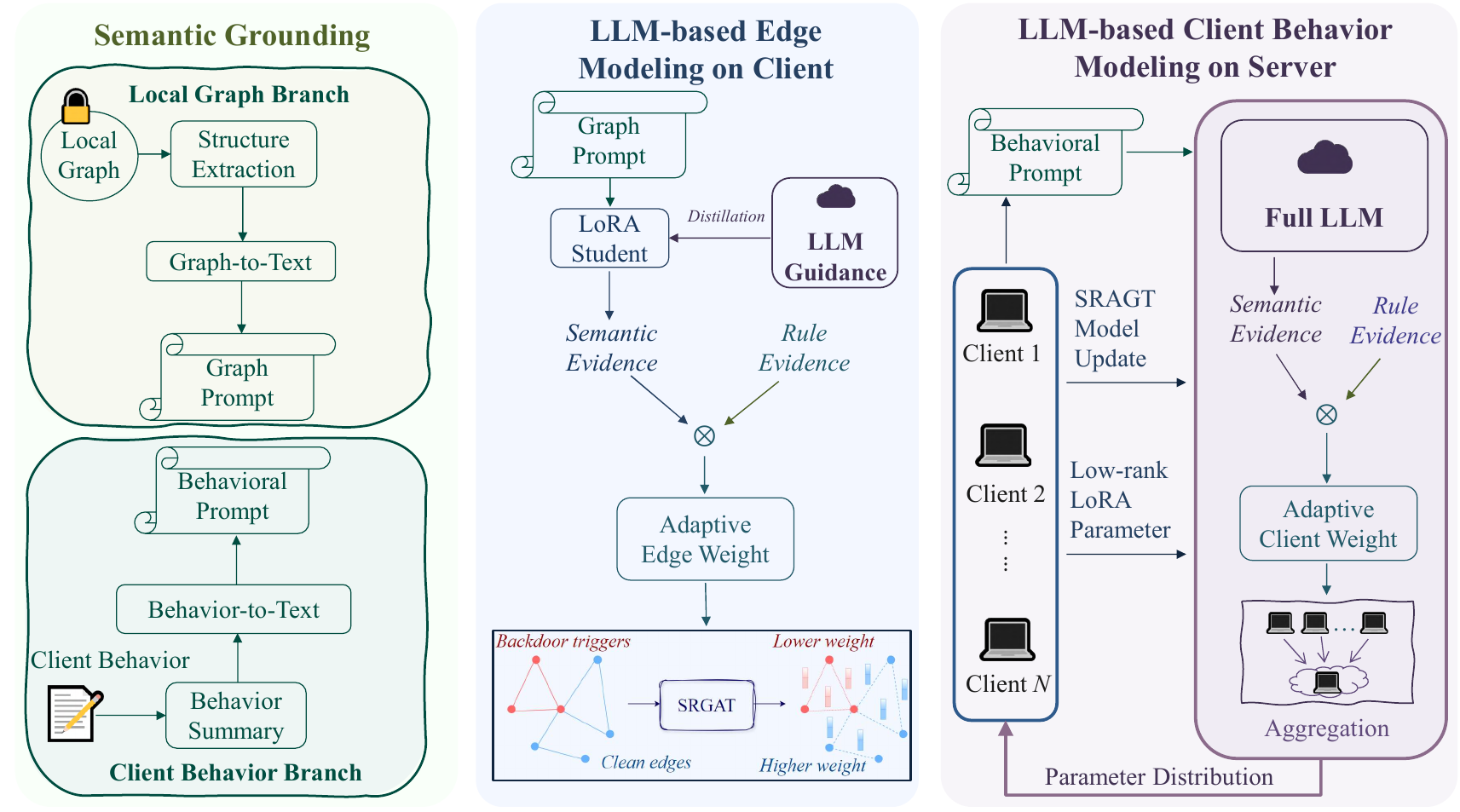}
\caption{FedLSG}
\label{Fig_wh}
\end{figure}

\section{Method}

We propose FedLSG, a training-time defense framework for mitigating backdoor attacks in FedGNNs. Moreover, the complexity analysis in Appendix 1 proves its efficiency.


\subsection{Backbone framework}

In FedGNN, some clients may be compromised by backdoor attacks, motivating the need to mitigate their effects and provide an implicit measure of client behavior, we design a simple yet robust graph attention model (SRGAT). Since backdoor triggers often inject task-specific edges that are difficult to distinguish from benign ones\cite{Yuan24sa,RIGBD,Yu25DS}, SRGAT applies a soft reweighting scheme that assigns attention scores to suppress suspicious connections.

Let $H_u$ denote the feature of node $u$, $\mathcal{N}(u)$ its neighbor set. For heterogeneous graphs, we use graphlet degree vectors instead of node features, following \cite{GNNGuard}. We build on GAT \cite{GAT} with attention weights that integrate structural and semantic signals using rules and LLM respectively, as follows:
\begin{equation}
\tilde{s}_{uv} = s_{uv} \cdot (Rule\_score + LLM\_score)  . \label{Eq_tsuv}
\end{equation}
\begin{equation}
\begin{aligned}
s_{uv} =
\frac{\exp (\sigma(a^{T}[W H_u\|W H_v]))}
{\sum_{k \in \mathcal{N}(u)}
\exp (\sigma(a^{T}[W H_u\|W H_k]))},
\end{aligned}
\label{Eq_sij}
\end{equation}
where $\sigma$ is the LeakyReLU activation, $a$ is a learnable parameter vector, $W$ is a weight matrix for node features, and $\|$ denotes concatenation. The attention weights $s_{uv}$ serve as an implicit indicator of edge reliability. The updated node representation is then computed as attention-weighted aggregation of neighbor features:
\begin{equation}
H_u^{\prime} = \sigma \left(\sum_{v \in\mathcal{N}(u)} \tilde{s}_{uv} W H_v \right) \label{Eq_Hi}.
\end{equation}

By integrating rule and LLM-derived scores into the attention mechanism, SRGAT adaptively downweights edges that are potentially associated with backdoor triggers, while preserving the original graph. We further decompose the framework into a server-side teacher and multiple client-side students, as follows:
\paragraph{Global Model:} The server consists of SRGAT and a full LLM, where SRGAT encodes graph structure, and the LLM provides semantic trust reasoning as the teacher.

\paragraph{Client Model:} Each client adopts SRGAT with a LoRA-based compensation, which distills knowledge from the teacher while preserving efficient local adaptation.

\subsection{LLM-based Edge Modeling on Client}

Each client in FedLSG maintains an SRGAT and a LoRA-based module to produce semantic evidence over structurally selected candidate edges. The resulting semantic signals is combined with the rule signal into SRGAT, to compute edge-level trust scores, which are used to suppress potentially backdoor edges during message passing rather than hard removal.

\paragraph{Rule-based edge scoring.} For each edge $(u,v)$ on client $i$, we first compute a structural anomaly score $r_{uv}^{\mathrm{rule}}$ (detailed in Appendix 2). This score is derived from local topology cues, including homophily, cross-community connectivity, neighborhood overlap, clustering consistency, and statistical anomaly indicators. Importantly, these signals are not interpreted as semantic judgments; instead, they serve as a coarse filtering mechanism to identify candidate edges that require deeper semantic inspection.

\paragraph{LLM-based edge scoring.}
To obtain semantic evaluations, FedLSG first constructs a graph-to-text reasoning pipeline with constrained neighborhood sampling, ensuring that identical graph structures always yield consistent textual evidence across clients.

We first define a instruction $\mathcal{S}^{\mathrm{det}}$. It specifies a fixed reasoning protocol: the model should interpret structural evidence around an edge and output a scalar suspiciousness score in a predefined format. Thus, $\mathcal{S}^{\mathrm{det}}$ determines how each edge is semantically analyzed in a consistent manner.

Given an edge $(u,v)$, we construct a prompt set $\mathcal{U}_{uv}^{\mathrm{edge}}$, where each prompt contains the same edge-level structural description together with  sampled neighbor context. The edge is first serialized into text, e.g.:

\emph{“Edge (u, v): node u and node v share common neighbor;  their community labels are different;  local clustering around u is high while v shows weak connectivity;  this edge connects two weakly related regions of the graph.”}

To avoid overloading the semantic model with full neighborhood information, we adopt a constrained neighbor querying strategy, where each prompt selectively samples high-degree neighbors and describes them, e.g.:

\emph{“Neighbor w: high-degree node with strong intra-community connectivity.”}

Thus, each element of $\mathcal{U}_{uv}^{\mathrm{edge}}$ integrates edge-level structure with partial neighborhood context, forming a multi-view yet bounded semantic representation.

Finally, a graph-derived semantic information is incorporated into the textual prompt representation as an auxiliary signal, and the final semantic score is computed from the combined input, as follows:
\begin{equation}
r_{uv}^{\mathrm{llm}} =
\mathcal{F}_{\mathrm{LLM}}^{\mathrm{edge}}
\!\left(\mathcal{S}^{\mathrm{det}}, \mathcal{U}_{uv}^{\mathrm{edge}}, \Delta_i^{\mathrm{LoRA}}\right),
\label{Eq_edge_llm}
\end{equation}
\begin{equation}
\Delta_i^{\mathrm{LoRA}} = A_iB_i, \qquad
\label{Eq_student_lora}
\end{equation}
where $\mathcal{F}_{\mathrm{LLM}}^{\mathrm{edge}}(\cdot,\cdot,\cdot)$ aggregates the $n_{uv}$ query scores by arithmetic mean. $A_i$ and $B_i$ are the client-specific LoRA adaptation matrices that parameterize the low-rank update.

\paragraph{Score fusion.} We then combine the rule score and the LLM score to form a unified edge risk estimate in SRGAT, as shown in Eq. \ref{Eq_sematt}, which serves to suppress potentially malicious edges.
\begin{equation}
\tilde{r}_{uv} = a_1^{edge} r_{uv}^{\mathrm{rule}} + a_2^{edge} r_{uv}^{\mathrm{llm}},
\label{Eq_rfuse}
\end{equation}
\begin{equation}
\tilde{s}_{uv} = s_{uv}(1-\kappa_e \tilde{r}_{uv}). \label{Eq_sematt}
\end{equation}

\subsection{Teacher–Student Knowledge Distillation}

To enable effective local optimization under limited client capacity, the client-side student is trained under supervision from the server-side teacher. The server broadcasts the current global SRGAT parameters $\theta^t$ together with the low-rank prior $(A^t, B^t)$, which jointly serve as shared structural and semantic references for all clients, as follows:
\begin{equation}
\mathcal{L}_{\mathrm{align}} = \left\|A_i-A^t\right\|_F^2 + \left\|B_i-B^t\right\|_F^2,
\label{Eq_align}
\end{equation}
which encourages alignment between local and global low-rank updates. In addition, we impose a low-rank regularizer as:
\begin{equation}
\mathcal{L}_{\mathrm{LoRA}} = \|A_i\|_F + \|B_i\|_F.
\label{Eq_lora}
\end{equation}

Beyond parameter-level alignment, we further transfer semantic and structural knowledge from the teacher to the student. Specifically, we perform distillation at three complementary levels: node-level representations, graph-level global semantics, and edge-level relational patterns:
\begin{equation}
\begin{aligned}
\mathcal{L}_{\mathrm{distill}} = &
\left\|Z_i^{\mathrm{stu}} - Z_i^{\mathrm{tea}}\right\|_2^2
+ \left\|\bar{Z}_i^{\mathrm{stu}} - \bar{Z}_i^{\mathrm{tea}}\right\|_2^2 \\
& + \sum_{(u,v)\in E_i}\left\|z_{uv}^{\mathrm{stu}} - z_{uv}^{\mathrm{tea}}\right\|_2^2,
\end{aligned}
\label{Eq_distill}
\end{equation}
where $Z_i$ denotes the representation obtained by student or teacher, $\bar{Z}_i$ denotes the graph-level mean representation, and $z_{uv}$ is the edge representation. Combining the cross-entropy loss $\mathcal{L}_{\mathrm{CE}}$ with the above losses, the following total loss can be obtained:
\begin{equation}
\mathcal{L} = \mathcal{L}_{\mathrm{CE}} + \lambda_1 \mathcal{L}_{\mathrm{align}} + \lambda_2 \mathcal{L}_{\mathrm{LoRA}} + \lambda_3 \mathcal{L}_{\mathrm{distill}}.
\label{Eq_splitloss}
\end{equation}

This objective ensures that the client-side student remains lightweight while preserving alignment with the server-side teacher.

\subsection{LLM-based Client Behavior Modeling on Server}

The server-side LLM not only provides global semantic priors to guide the client-side student, but also performs semantic reasoning over client behavioral signals during the aggregation stage, enabling the identification of potentially malicious clients and the adaptive down-weighting of their contributions.

\paragraph{Rule-based client scoring.} We construct a client-level rule trust score $\tau_{i,\mathrm{client}}^{\mathrm{rule}}$ based on structural summary statistics, including homophily deviation, anomaly indicators, and attention-derived structural cues, detailed in Appendix 2.

\paragraph{LLM-based client scoring.} We define a fixed instruction template $\mathcal{S}^{\mathrm{client}}$, which guides the semantic student to interpret client behavior under a unified reasoning protocol and output a scalar behavioral trust score. This ensures that all clients are assessed under consistent semantic criteria.

Meanwhile, $\mathcal{U}^{\mathrm{client}}$ is the client-level semantic input, to represent multi-view client behavior, consisting of:
(i) structural behavioral summaries capturing graph-level interaction patterns and anomaly tendencies; (ii) cohort-normalized behavioral statistics reflecting deviations from benign population behavior; (iii) student embeddings that encode behavior-aware semantic representations distilled from the server-side teacher, including how the client reacts to global knowledge.

To illustrate the construction process, a prompt instance is formed by instantiating client behaviors, for example:

\emph{"Evaluate client trustworthiness using its structural summary, cohort-normalized statistics, and student embedding, where the client shows moderate homophily deviation and anomaly signals relative to cohort baselines; output a scalar score indicating whether it is benign or abnormal."}

Then, these prompts are uesd to produce the behavioral semantic trust score
$\tau_{i,\mathrm{client}}^{\mathrm{llm}}$, as follows:
\begin{equation}
\tau_{i,\mathrm{client}}^{\mathrm{llm}} =
\mathcal{F}_{\mathrm{llm}}^{\mathrm{client}}
\!\left(\mathcal{S}^{\mathrm{client}}, \mathcal{U}^{\mathrm{client}}, i\right),
\label{Eq_client_llm}
\end{equation}
where $\mathcal{F}_{\mathrm{llm}}^{\mathrm{client}}(\cdot)$ produces a scalar client behavioral trust score under LLM reasoning.

\paragraph{Aggregation weight.} We then combine behavioral rule trust and semantic trust to obtain a unified estimate:
\begin{equation}
\tilde{\tau}_{i,\mathrm{client}} = a_{1}^\mathrm{client} \tau_{i,\mathrm{client}}^{\mathrm{rule}} + a_{2}^\mathrm{client} \tau_{i,\mathrm{client}}^{\mathrm{llm}} 
\label{Eq_clientfuse}
\end{equation}

The final aggregation weight is:
\begin{equation}
w_i^t = 1-\kappa_a (1-\tilde{\tau}_{i,\mathrm{client}})\omega_t. \label{Eq_aggw}
\end{equation}

Here $\kappa_a\in(0,1]$ controls the maximum trust-based aggregation correction and $w_i^t\in[1-\kappa_a,1]$ is the unnormalized aggregation weight assigned to client $i$ at round $t$. 

For FedAvg-style aggregation, the server updates:
\begin{equation}
\begin{aligned}
\theta^{t+1} ={}& \sum_{i=1}^{N_c}\bar{w}_i^t\theta_i^t, \\
\bar{w}_i^t ={}& \frac{w_i^t}{\sum_{j=1}^{N_c}w_j^t}.
\end{aligned}
\label{Eq_weightedavg}
\end{equation}
where $\bar{w}_i^t$ is the normalized aggregation coefficient and satisfies $\sum_{i=1}^{N_c}\bar{w}_i^t=1$. This behavioral trust mechanism prevents clients exhibiting abnormal training or semantic behaviors from dominating the global model, even if partial poisoned signals survive local filtering.

Since full semantic reasoning is not deployed on clients, only SRGAT parameters and low-rank behavioral adaptation parameters $(A_i^t, B_i^t)$ are communicated to the server. The server aggregates these parameters as:
\begin{equation}
\begin{aligned}
A^{t+1} ={}& \sum_{i=1}^{N_c}\bar{w}_i^t A_i^t, \\
B^{t+1} ={}& \sum_{i=1}^{N_c}\bar{w}_i^t B_i^t.
\end{aligned}
\label{Eq_loraagg}
\end{equation}
The aggregated $(A^{t+1},B^{t+1})$ define the next-round global low-rank prior, which is redistributed to clients together with the global SRGAT parameters $\theta^{t+1}$. In this way, aggregation remains unchanged in form, while being guided by client behavioral semantics learned from both rule-based and LLM-based evidence. 

\subsection{Rule-LLM Trust Estimation}

Although rule structural indicators efficiently detect suspicious patterns, they are sensitive to benign non-IID variations and may misclassify normal deviations. Split-semantic reasoning provides higher-level context but is limited by truncated semantic depth.  Thus, neither is fully reliable in isolation. FedLSG introduces an adaptive mechanism that balances both evidence sources for trust estimation.

Specifically, trust is evaluated from two evidences: rule evidence and LLM evidence, computed from Eq. \ref{Eq_rfuse} and \ref{Eq_clientfuse}, respectively. For each evidence $q$, the rule module outputs $\tau_q^{rule}$ and the LLM module outputs $\tau_q^{llm}$. And a disagreement factor $D_q$ measures their inconsistency, where larger values indicate stronger divergence. Based on $D_q$, two utility functions $M_1$ and $M_2$ are defined to quantify the confidence of rule and LLM evidence, respectively.
\begin{gather}
\begin{cases}
\begin{aligned}
&M_{1}^{q}(rule, llm)=\sigma_s\left(\tau_{q}^{rule}\cdot(1-D_{q})\right),\\
&M_{2}^{q}(rule, llm)=\sigma_s\left(\tau_{q}^{llm}\cdot(1-D_{q})\right),
\end{aligned}
\end{cases}
\label{Eq_M1M2_}
\end{gather}

\begin{equation}
\begin{aligned}
D_{q} = max \left( |\tau_{q}^{rule} - \tau_{q}^{llm}|, \epsilon  \right)
\end{aligned}
\label{Eq_Dq}
\end{equation}
where $\sigma_s(\cdot)$ denotes the sigmoid function and $\epsilon$ is a small constant for numerical stability. When the rule and LLM assessments are highly consistent, $D_{trust}^{q}$ becomes small, leading to larger confidence values. Conversely, strong disagreement reduces both utilities and discourages over-reliance on potentially unreliable evidence.

Motivated by prior adaptive interaction modeling \cite{Lu20GT,Liu18GT}, FedLSG formulates their interaction as a optimization problem:
\begin{gather}
\begin{aligned}
& \operatorname*{min} f = \sum_{q=1}^{Q} \left| \left( \sum_{p=1}^{P} a_p M_p^q(rule, llm) \right) - M_p^q(rule, llm) \right| \\
& \qquad \qquad \mathrm{s.t.} \quad a_1, a_2 > 0; \frac{\lambda}{P} \left( \sum_{p=1}^{P} a_p^2 \right) = \gamma,
\end{aligned}
\label{Eq_minf_}
\end{gather}
where ($a_1$) and $a_2$ denote the adaptive weights assigned to rule and LLM evidence, respectively.And $\gamma$ is a constraint coefficient controlling the solution space. $P=2$ denotes the two competing objectives, i.e., rule and LLM assessment.  $Q$ represents the number of evidences, corresponding to edge-level and client-level  evidence.  In our framework, $Q$ is stage-dependent: edge-level estimation (Eq. \ref{Eq_rfuse}) uses $Q=1$, while client-level estimation (Eq. \ref{Eq_clientfuse}) incorporates cross-client evidence with $Q=2$.

Solving Eq.~\ref{Eq_minf_} yields the optimal contribution of each evidence source:

\begin{equation}
a_p =
\frac{\sqrt{\gamma P}\cdot \sum_{q=1}^{Q} M_p^q(rule, llm)}
{\sqrt{\sum_{p=1}^{P}\left(\sum_{q=1}^{Q} M_p^q(rule, llm)\right)^2}},
\label{Eq_ab}
\end{equation}
where larger utility values naturally result in greater influence on the final trust estimation. The calibrated client trust score is subsequently used to guide trust-aware aggregation and suppress the backdoor patterns.




\section{Experiment}

We conduct a series of experiments to evaluate the effectiveness of FedLSG. Additional results, including LLM backbone selection, ablation studies, parameter analysis, heterogeneous graphs, varying numbers of attacks and clients, and IID settings, are provided in Appendix 6-12.

\begin{table}[h]
\centering
\caption{Data statistics}
\label{Tab_data}
\begin{tabular}{lcccc}
\toprule
Datasets   & Nodes  & Edges   & Classes & Features \\ \hline
Citeseer   & 3327   & 9104    & 6       & 3703     \\
Pubmed     & 19717  & 44338   & 3       & 500      \\
Flickr     & 89250  & 899756  & 7       & 500      \\
Ogb-arxiv & 169343 & 1166243 & 40      & 128     \\
\bottomrule
\end{tabular}
\end{table}

\textbf{Datasets:} We test the effectiveness of FedLSG on several publicly available datasets, including Citeseer \cite{Sen08P}, Pubmed \cite{Sen08P}, Flickr \cite{Flickr}, Ogb-arxiv \cite{Ogbn}. Among these, Citeseer and Pubmed are small datasets, While Flickr and Ogb-arxiv are large-scale datasets \cite{UGBA}. The statistics are provided in Table \ref{Tab_data}. We follow the setup of \cite{Bkdfedgnn}, use the non-identically distributed Louvain (Non-IID-Louvain) data partition setting.

\textbf{Attacks:} We adopt the state-of-the-art graph backdoor attacks DPGBA \cite{DPGBA}, UGBA \cite{UGBA}, and SBA \cite{SBA}, to test the defense of FedLSG. For each attack, we use the attack parameters suggested by the authors. Furthermore, we follow the setup of \cite{Bkdfedgnn} to adapt the above attacks to the graph federation system. We randomly attack 60 percent of clients. We run 200 rounds, performing attacks in each rounds.

\textbf{Baselines:} We test state-of-the-art defense methods including Prune \cite{UGBA}, PruneLD \cite{UGBA}, OD \cite{DPGBA}, RLR \cite{RLR}, NoisyGCN \cite{NoisyGCN}, FedTGE \cite{FedTGE}, RIGBD \cite{RIGBD}, and the no-defense GCN (None). These methods are described in Appendix 5. And we evaluate the attack success rate (ASR) when backdoor attacks target the above methods, as well as their classification accuracy (ACC) on clean data. A lower ASR and a higher ACC reflect better defense performance.

\textbf{Parameters:} FedLSG is trained with 20 clients, hidden dimensions to $\{128,256\}$, learning rate to 0.005, weight decay to $\{0,5\times10^{-4}\}$, dropout rate to $\{0.1,0.5\}$, and $\gamma=1$. Adam is the optimization, with Gemma-4-12B as the LLM backbone \cite{llama24}. Experiments are conducted on four NVIDIA RTX 4090 GPUs. Detailed settings and LLM comparisons are reported in Appendix 6 and 7.

\subsection{Defense Effectiveness Evaluation} 

In this section, we compare the defensive efficacy of FedLSG and baselines in defending against DPGBA, UGBA, and SBA on Citeseer, Pubmed, Flickr, and Ogb-arxiv. Initially, we report the results as shown in Table \ref{Tab_def}. The lowest ASR values are highlighted in bold.

\begin{table}[h]
\centering
\caption{The results (ASR and ACC) between FedLSG and baselines}
\scriptsize  
\begin{tabular}{llcccccccc}
\toprule
   &  & \multicolumn{2}{c}{Citeseer}   &\multicolumn{2}{c}{Pubmed}   &\multicolumn{2}{c}{Flickr}   &\multicolumn{2}{c}{Ogb-arxiv}   \\ \cline{3-10}
\multirow{-2}{*}{} & \multirow{-2}{*}{Defenses}   &ASR$\downarrow$  &ACC$\uparrow$  &ASR$\downarrow$  &ACC$\uparrow$  &ASR$\downarrow$  &ACC$\uparrow$  &ASR$\downarrow$  &ACC$\uparrow$  \\ \hline
&None   & 92.37$\pm$0.95                                 & 74.76$\pm$1.15                                  & 93.46$\pm$0.86                                  & 85.72$\pm$0.95                                  & 87.19$\pm$1.42                                 & 46.80$\pm$1.15                                  & 95.35$\pm$1.27                                  & 65.67$\pm$0.27    \\
&Prune  & 82.58$\pm$3.16                                 & 72.90$\pm$0.35                                  & 46.33$\pm$0.99                                  & 81.74$\pm$1.12                                  & 54.35$\pm$1.08                                 & 46.39$\pm$0.68                                  & 60.07$\pm$4.24                                  & 62.36$\pm$0.42    \\
&PruneLD  & 82.28$\pm$3.20                                 & 72.66$\pm$1.91                                  & 45.47$\pm$0.96                                  & 81.13$\pm$0.68                                  & 54.08$\pm$1.26                                 & 46.45$\pm$1.05                                  & 56.79$\pm$5.34                                  & 62.39$\pm$1.04    \\
&OD  & 91.85$\pm$0.36                                 & 74.08$\pm$1.74                                  & 41.95$\pm$0.81                                  & 83.54$\pm$0.99                                  & 38.21$\pm$1.93                                 & 46.34$\pm$0.56                                  & 47.63$\pm$2.26                                  & 61.51$\pm$1.07    \\
&RLR    & 63.02$\pm$2.55                                 & 72.29$\pm$1.02                                  & 40.51$\pm$1.31                                  & 85.53$\pm$1.22                                  & 36.55$\pm$2.60                                 & 46.01$\pm$0.91                                  & 42.32$\pm$1.01                                  & 66.30$\pm$0.98    \\
&NoisyGCN   & 54.54$\pm$1.17                                 & 73.12$\pm$1.91                                  & 25.08$\pm$0.99                                  & 82.12$\pm$0.88                                  & 35.02$\pm$2.40                                 & 45.71$\pm$1.76                                  & 40.75$\pm$3.19                                  & 65.31$\pm$0.87    \\
&FedTGE   & 35.47$\pm$2.76                                 & 74.51$\pm$0.61                                  & 21.23$\pm$1.70                                  & 84.87$\pm$1.24                                  & 28.34$\pm$1.28                                 & 44.87$\pm$0.47                                  & 46.32$\pm$2.97                                  & 65.11$\pm$0.50 \\
&RIGBD   & 21.52$\pm$1.89                                 & 70.78$\pm$0.90                                  & 20.97$\pm$0.20                                  & 81.08$\pm$1.71                                  & 26.25$\pm$0.87                                 & 44.26$\pm$1.37                                  & 42.79$\pm$2.73                                  & 61.98$\pm$0.99    \\
\multirow{-8}{*}{\rotatebox{90}{DPGBA}}   &\textbf{FedLSG} &\textbf{5.58$\pm$0.82} &\textbf{74.85$\pm$0.59} &\textbf{7.61$\pm$0.66} &\textbf{86.14$\pm$0.31} &\textbf{4.52$\pm$0.55} &\textbf{45.70$\pm$0.58} &\textbf{30.21$\pm$0.68} &\textbf{66.22$\pm$0.78} \\  
\hline
&None   & 99.67$\pm$1.37                                 & 74.60$\pm$0.56                                  & 94.19$\pm$0.33                                  & 84.81$\pm$0.71                                  & 83.26$\pm$1.71                                 & 45.10$\pm$1.05                                  & 99.54$\pm$0.96                                  & 65.79$\pm$0.98   \\
&Prune   & 55.86$\pm$1.48                                 & 72.50$\pm$0.40                                  & 89.80$\pm$4.06                                  & 83.03$\pm$0.64                                  & 54.84$\pm$4.18                                 & 45.26$\pm$1.06                                  & 89.00$\pm$3.36                                  & 62.16$\pm$1.11    \\
&PruneLD   & 54.05$\pm$1.50                                 & 71.87$\pm$1.05                                  & 88.34$\pm$4.04                                  & 83.23$\pm$0.46                                  & 56.59$\pm$3.22                                 & 45.42$\pm$0.81                                  & 86.25$\pm$2.03                                  & 62.40$\pm$1.16    \\
&OD  & 40.57$\pm$0.28                                 & 75.34$\pm$0.88                                  & 80.71$\pm$3.15                                  & 84.55$\pm$0.08                                  & 0.00$\pm$0.00                                  & 45.26$\pm$2.21                                  & 73.47$\pm$4.15                                  & 65.45$\pm$0.93    \\
&RLR    & 97.58$\pm$1.12                                 & 73.60$\pm$1.13                                  & 78.75$\pm$1.06                                  & 85.29$\pm$0.38                                  & 78.57$\pm$2.39                                 & 45.71$\pm$1.18                                  & 1.78$\pm$0.47                                   & 65.89$\pm$0.79    \\
&NoisyGCN  & 62.49$\pm$0.76                                 & 55.21$\pm$1.46                                  & 64.13$\pm$1.99                                  & 78.53$\pm$0.29                                  & 0.00$\pm$0.00                                  & 45.80$\pm$1.22                                  & 1.22$\pm$0.38                                   & 61.74$\pm$1.05    \\
&FedTGE & 19.67$\pm$1.80                                 & 73.59$\pm$1.13                                  & 77.39$\pm$2.31                                  & 85.08$\pm$0.93                                  & 69.79$\pm$3.12                                 & 44.86$\pm$0.37                                  & 72.09$\pm$3.98                                  & 62.57$\pm$0.46 \\
&RIGBD   & 15.68$\pm$3.92                                 & 70.58$\pm$0.89                                  & 59.36$\pm$0.27                                  & 81.04$\pm$0.79                                  & 0.00$\pm$0.00                                  & 42.50$\pm$0.89                                  & 7.58$\pm$0.92                                   & 63.14$\pm$0.73    \\
\multirow{-8}{*}{\rotatebox{90}{UGBA}}   &\textbf{FedLSG} &\textbf{2.75$\pm$0.56} &\textbf{75.45$\pm$0.42} &\textbf{32.30$\pm$0.78} &\textbf{85.13$\pm$0.43} &\textbf{0.00$\pm$0.00} &\textbf{46.29$\pm$0.53} &\textbf{0.36$\pm$0.21}  &\textbf{65.50$\pm$0.61} \\  
\hline
&None   & 47.82$\pm$1.21                                 & 73.49$\pm$1.07                                  & 50.97$\pm$1.15                                  & 85.13$\pm$0.84                                  & 0.00$\pm$0.00                                  & 45.39$\pm$1.31                                  & 0.00$\pm$0.00                                   & 64.44$\pm$1.82    \\
&Prune  & 9.31$\pm$1.19                                  & 70.73$\pm$1.21                                  & 19.47$\pm$6.14                                  & 81.47$\pm$0.92                                  & 0.00$\pm$0.00                                  & 44.55$\pm$0.89                                  & 0.00$\pm$0.00                                   & 64.33$\pm$1.13    \\
&PruneLD   & 8.41$\pm$1.50                                  & 70.52$\pm$0.27                                  & 18.81$\pm$5.01                                  & 81.76$\pm$0.90                                  & 0.00$\pm$0.00                                  & 44.35$\pm$1.47                                  & 0.00$\pm$0.00                                   & 64.60$\pm$0.74    \\
&OD    & 17.42$\pm$1.99                                 & 71.49$\pm$1.79                                  & 21.04$\pm$5.76                                  & 83.39$\pm$0.42                                  & 0.00$\pm$0.00                                  & 45.71$\pm$0.66                                  & 0.00$\pm$0.00                                   & 64.57$\pm$0.69    \\
&RLR    & 15.31$\pm$2.65                                 & 73.72$\pm$0.97                                  & 20.79$\pm$3.47                                  & 84.71$\pm$0.11                                  & 0.00$\pm$0.00                                  & 45.48$\pm$1.04                                  & 0.00$\pm$0.00                                   & 64.03$\pm$1.69    \\
&NoisyGCN  & 7.60$\pm$0.53                                  & 66.82$\pm$0.93                                  & 11.95$\pm$1.29                                  & 81.44$\pm$0.09                                  & 0.00$\pm$0.00                                  & 44.59$\pm$1.08                                  & 0.00$\pm$0.00                                   & 64.42$\pm$1.70    \\
&FedTGE   & 12.58$\pm$2.68                                 & 73.59$\pm$0.52                                  & 18.78$\pm$2.25                                  & 84.90$\pm$0.70                                  & 0.00$\pm$0.00                                  & 44.03$\pm$0.81                                  & 0.00$\pm$0.00                                   & 64.36$\pm$1.04 \\
&RIGBD   & 4.70$\pm$0.65                                  & 70.18$\pm$0.71                                  & 10.24$\pm$4.47                                  & 81.33$\pm$0.95                                  & 0.00$\pm$0.00                                  & 44.91$\pm$0.53                                  & 0.00$\pm$0.00                                   & 63.81$\pm$0.74    \\ 
\multirow{-8}{*}{\rotatebox{90}{SBA}}    &\textbf{FedLSG} &\textbf{0.00$\pm$0.00} &\textbf{74.97$\pm$0.49} &\textbf{0.13$\pm$0.39}  &\textbf{85.45$\pm$0.58} &\textbf{0.00$\pm$0.00} &\textbf{46.25$\pm$0.37} &\textbf{0.00$\pm$0.00}  &\textbf{64.90$\pm$0.51}  \\ \bottomrule
\end{tabular}
\label{Tab_def}
\end{table}

Experimental results show that FedLSG achieves the lowest ASR across all datasets and all settings, while preserving clean-data performance. DPGBA and UGBA are very strong attacks, maintaining high ASRs in the undefended case, while SBA is weaker, with ASRs consistently at 0 in the extra-large datasets Flickr and Ogb-arxiv, even when the attack strength is increased. This phenomenon is not due to reporting errors, but rather stems from the limited capability of SBA in large-scale graph settings. Specifically, SBA relies on injecting relatively simple and localized backdoor triggers, which become less effective when the graph size increases significantly. In large graphs such as Flickr and Ogb-arxiv, the influence of a small number of poisoned nodes is diluted during message passing and aggregation, preventing the backdoor from significantly influencing model outputs. Moreover, UGBA achieves remarkable attack performance on Pubmed, leading to an attack success rate even higher than that on larger datasets. This may be because Pubmed exhibits stronger class homophily and semantic consistency, allowing the injected backdoor patterns to align more easily with the target class, thereby improving both the stealthiness and effectiveness of the attack.

Regarding the defense methods, Prune, PruneLD, OD hard to defend against DPGBA, which generates intra-distribution triggers \cite{RIGBD}. RLR and FedTGE are used to defend in the client aggregation phase, but lack the ability to protect clean clients. NoisyGCN and RIGBD provide stronger defenses but sacrifice too much clean ACC. Additionally, all the above baselines share a common issue: in federated systems, it is unknown which clients are attacked or when. As a result, these baselines must be applied to each client, which inevitably reduces the clean accuracy, especially when the percentage of pruned edges is high. Furthermore, no defense can completely eliminate backdoors, as a few poisoned target nodes may behave similarly to clean nodes \cite{RIGBD}, inevitably remaining in the graph and disrupting the client model. This can affect the central model during aggregation, which, in turn, impacts other clean clients in the next round, leading to a cascading failure \cite{Pmlr22zh,CAI24FL,NGU24Ba}. In contrast, FedLSG adaptively adjusts the heterogeneity between clients, to mitigate the impact of backdoors to the central model, thereby providing effective defense.

\subsection{Defense in white-box adaptive attack scenarios}

We evaluate the defense of FedLSG in a white-box setting where the attacker has full knowledge of the LLM, SRGAT and adaptive mechanism. The adaptive attacks DPGBA and UGBA are adapted to optimize perturbations with respect to the entire FedLSG framework, rather than treating the defense as a fixed preprocessing step. Concretely, FedLSG is used as the surrogate model, enabling gradient-based attacks to explicitly target the LLM, SRGAT and adaptive mechanism. This adaptive white-box setting reflects a worst-case threat model. The results in Figure \ref{Fig_wh} demonstrate that FedLSG provides strong defense against both centralized and distributed adaptive white-box attacks, despite a slight increase in ASR. Notably, FedLSG under adaptive white-box attacks even outperforms most baseline methods evaluated in non-white-box settings.

\begin{figure}[h]
\centering
\includegraphics[width=0.60\columnwidth]{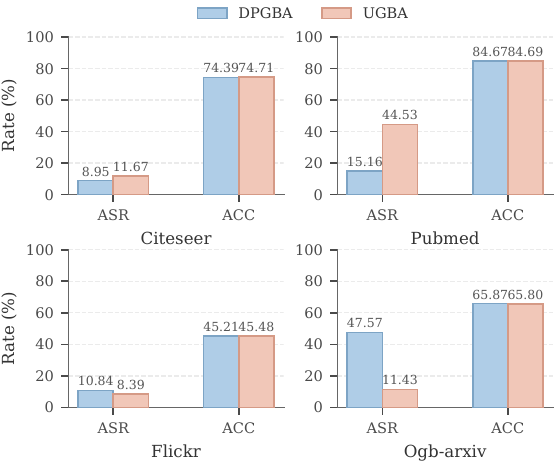}
\caption{Defense in white-box adaptive attack scenarios}
\label{Fig_wh}
\end{figure}

\subsection{Impact of the number of attacks}

\begin{figure}[h]
\centering
\includegraphics[width=0.6\columnwidth]{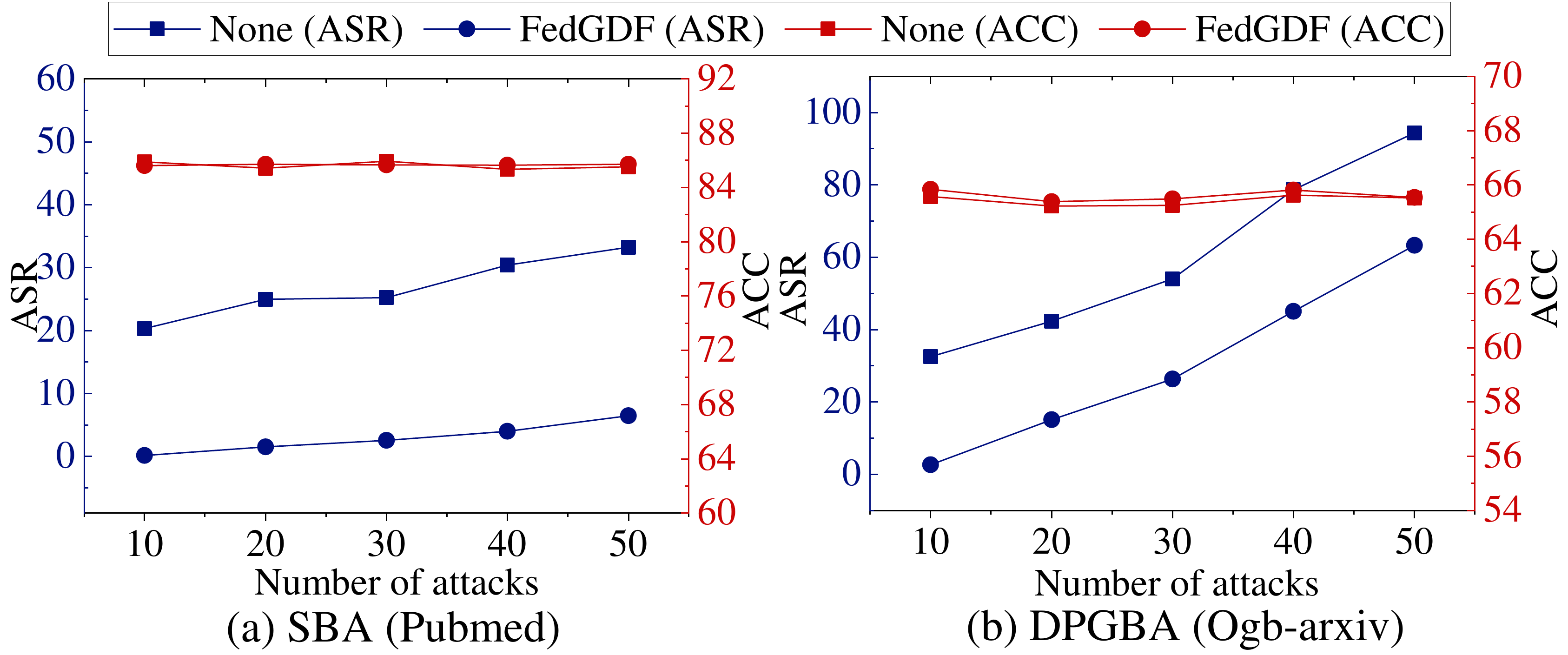}
\caption{Impact of the number of attacks}
\label{Fig_NA}
\end{figure}

We explore the impact of the number of attacked clients on FedLSG, as presented in Figure \ref{Fig_NA}. Specifically, we vary the number of attacked clients from 10 to 50, with a total of 50 clients. We compare the No-defense and FedLSG against SBA and DPGBA attacks on Pubmed and Ogb-arxiv datasets. As shown in Figure \ref{Fig_NA}, the ASR generally increases with the number of compromised clients in No-defense condition, which is consistent with the intuitive expectation that more malicious participants introduce stronger interference into the global model. For comparison, FedLSG exhibits a remarkably stable and low ASR across all levels of attack intensity, indicating its strong resistance to severe attacks.

\section*{Acknowledgments}
We sincerely thank all anonymous reviewers for their valuable efforts.

\bibliographystyle{unsrt}  
\bibliography{references}

@article{Sen08P,
author = {Sen, Prithviraj and Namata, Galileo and Bilgic, Mustafa and Getoor, Lise and Gallagher, Brian and Eliassi-Rad, Tina},
journal = {AI Magazine},
number = {3},
pages = {93--106},
title = {Collective Classification in Network Data},
volume = {29},
year = {2008}
}

@inproceedings{UGBA,
author = {Dai, Enyan and Lin, Minhua and Zhang, Xiang and Wang, Suhang},
booktitle = {Proceedings of the ACM Web Conference},
pages = {2263--2273},
title = {Unnoticeable Backdoor Attacks on Graph Neural Networks},
year = {2023}
}

@inproceedings{SBA,
author = {Zhang, Zaixi and Jia, Jinyuan and Wang, Binghui and Gong, Neil Zhenqiang},
pages = {15--26},
title = {Backdoor Attacks to Graph Neural Networks},
booktitle = {Proceedings of the ACM Symposium on Access Control Models and Technologies},
year = {2021}
}

@inproceedings{DPGBA,
author = {Zhang, Zhiwei and Lin, Minhua and Dai, Enyan and Wang, Suhang},
title = {Rethinking Graph Backdoor Attacks: A Distribution-Preserving Perspective},
year = {2024},
doi = {10.1145/3637528.3671910},
booktitle = {Proceedings of the ACM SIGKDD Conference on Knowledge Discovery and Data Mining},
pages = {4386–4397},
numpages = {12}
}

@inproceedings{NoisyGCN,
author = {Ennadir, Sofiane and Abbahaddou, Yassine and Lutzeyer, Johannes F and Vazirgiannis, Michalis and Bostr{\"{o}}m, Henrik},
booktitle = {Proceedings of the AAAI Conference on Artificial Intelligence},
pages = {21063--21071},
title = {A Simple and Yet Fairly Effective Defense for Graph Neural Networks},
year = {2024}
}

@article{Bkdfedgnn,
  title={Bkd-fedgnn: A benchmark for classification backdoor attacks on federated graph neural network},
  author={Liu, Fan and Lai, Siqi and Ning, Yansong and Liu, Hao},
  journal={arXiv preprint arXiv:2306.10351},
  year={2023}
}

@inproceedings{Mostly,
author = {Xu, Jing and Wang, Rui and Koffas, Stefanos and Liang, Kaitai and Picek, Stjepan},
title = {More is Better (Mostly): On the Backdoor Attacks in Federated Graph Neural Networks},
year = {2022},
doi = {10.1145/3564625.3567999},
booktitle = {Proceedings of the Annual Computer Security Applications Conference},
pages = {684–698},
numpages = {15}
}

@inproceedings{Flickr,
title={GraphSAINT: Graph Sampling Based Inductive Learning Method},
author={Hanqing Zeng and Hongkuan Zhou and Ajitesh Srivastava and Rajgopal Kannan and Viktor Prasanna},
booktitle={Proceedings of the International Conference on Learning Representations},
year={2020}
}

@inproceedings{Ogbn,
 author = {Hu, Weihua and Fey, Matthias and Zitnik, Marinka and Dong, Yuxiao and Ren, Hongyu and Liu, Bowen and Catasta, Michele and Leskovec, Jure},
 booktitle = {Proceedings of the Advances in Neural Information Processing Systems},
 pages = {22118--22133},
 title = {Open Graph Benchmark: Datasets for Machine Learning on Graphs},
 volume = {33},
 year = {2020}
}

@inproceedings{RIGBD,
title={Robustness Inspired Graph Backdoor Defense},
author={Zhiwei Zhang and Minhua Lin and Junjie Xu and Zongyu Wu and Enyan Dai and Suhang Wang},
booktitle={Proceedings of the International Conference on Learning Representations},
year={2025}
}

@inproceedings{RLR,
  title={Defending against backdoors in federated learning with robust learning rate},
  author={Ozdayi, Mustafa Safa and Murat, Kantarcioglu and Gel, Yulia R},
  booktitle={Proceedings of the AAAI conference on artificial intelligence},
  volume={35},
  number={10},
  pages={9268--9276},
  year={2021}
}

@inproceedings{lv25dy,
  title={Dynamic Multi-Interest Graph Neural Network for Session-Based Recommendation},
  author={Lv, Mingyang and Liu, Xiangfeng and Xu, Yuanbo},
  booktitle={Proceedings of the AAAI Conference on Artificial Intelligence},
  volume={39},
  number={12},
  pages={12328--12336},
  year={2025}
}

@article{Li1056,
  author={Li, Junwei and Wu, Le and Du, Yulu and Hong, Richang and Li, Weisheng},
  journal={IEEE Transactions on Computational Social Systems}, 
  title={Dual Graph Neural Networks for Dynamic Users' Behavior Prediction on Social Networking Services}, 
  year={2024},
  volume={11},
  number={5},
  pages={7020-7031},
  doi={10.1109/TCSS.2024.3409383}
}

@inproceedings{zhu25re,
  title={Refine then Classify: Robust Graph Neural Networks with Reliable Neighborhood Contrastive Refinement},
  author={Zhuang, Shuman and Wu, Zhihao and Chen, Zhaoliang and Dai, Hong-Ning and Liu, Ximeng},
  booktitle={Proceedings of the AAAI Conference on Artificial Intelligence},
  volume={39},
  number={12},
  pages={13473--13482},
  year={2025}
}

@inproceedings{Pmlr22zh,
title = {Neurotoxin: Durable Backdoors in Federated Learning},
author = {Zhang, Zhengming and Panda, Ashwinee and Song, Linyue and Yang, Yaoqing and Mahoney, Michael and Mittal, Prateek and Kannan, Ramchandran and Gonzalez, Joseph},
booktitle = {Proceedings of the International Conference on Machine Learning},
pages = {26429--26446},
year = {2022},
volume = {162}
}

@article{CAI24FL,
title = {FLMAAcBD: Defending against backdoors in Federated Learning via Model Anomalous Activation Behavior Detection},
author = {Hongyun Cai and Jiahao Wang and Lijing Gao and Fengyu Li},
journal = {Knowledge-Based Systems},
volume = {289},
pages = {111511},
year = {2024},
doi = {/10.1016/j.knosys.2024.111511}
}

@article{NGU24Ba,
title = {Backdoor attacks and defenses in federated learning: Survey, challenges and future research directions},
journal = {Engineering Applications of Artificial Intelligence},
volume = {127},
pages = {107166},
year = {2024},
doi = {/10.1016/j.engappai.2023.107166},
author = {Thuy Dung Nguyen and Tuan Nguyen and Phi Le Nguyen and Hieu H. Pham and Khoa D. Doan and Kok-Seng Wong}
}

@inproceedings{Wa22Fg,
author = {Wang, Zhen and Kuang, Weirui and Xie, Yuexiang and Yao, Liuyi and Li, Yaliang and Ding, Bolin and Zhou, Jingren},
title = {FederatedScope-GNN: Towards a Unified, Comprehensive and Efficient Package for Federated Graph Learning},
year = {2022},
doi = {10.1145/3534678.3539112},
booktitle = {Proceedings of the ACM SIGKDD Conference on Knowledge Discovery and Data Mining},
pages = {4110–4120},
numpages = {11}
}

@inproceedings{Yuan24sa,
  title={E-SAGE: Explainability-Based Defense Against Backdoor Attacks on Graph Neural Networks},
  author={Yuan, Dingqiang and Xu, Xiaohua and Yu, Lei and Han, Tongchang and Li, Rongchang and Han, Meng},
  booktitle={Proceedings of the International Conference on Wireless Artificial Intelligent Computing Systems and Applications},
  pages={402--414},
  year={2024}
}

@inproceedings{Yu25DS,
  title={DShield: Defending against Backdoor Attacks on Graph Neural Networks via Discrepancy Learning},
  author={Yu, Hao and Ma, Chuan and Wan, Xinhang and Wang, Jun and Xiang, Tao and Shen, Meng and Liu, Xinwang},
  booktitle={Proceedings of the Network and Distributed System Security Symposium, NDSS},
  year={2025}
}

@article{Lu20GT,
author = {Dan, Lu and Changqing, Xu and Linjuan, Zhang and Lili, Wang and Yanying, Sun},
doi = {10.1088/1755-1315/453/1/012068},
journal = {IOP Conference Series: Earth and Environmental Science},
number = {1},
pages = {12068},
title = {Comprehensive Risk Assessment Method of Power Grid Based on Grey Relational Weight Game Theory},
volume = {453},
year = {2020}
}

@article{Liu18GT,
author = {Liu, Tianyu and Deng, Yong and Chan, Felix T S},
doi = {10.1007/S40815-017-0400-4},
journal = {International Journal of Fuzzy Systems},
number = {4},
pages = {1321--1333},
title = {Evidential Supplier Selection Based on {DEMATEL} and Game Theory},
volume = {20},
year = {2018}
}

@inproceedings{Shen24Be,
 author = {Shen, Zhixiang and Wang, Shuo and Kang, Zhao},
 booktitle = {Proceedings of the Advances in Neural Information Processing Systems},
 pages = {31629--31658},
 title = {Beyond Redundancy: Information-aware Unsupervised Multiplex Graph Structure Learning},
 volume = {37},
 year = {2024}
}

@inproceedings{FedTGE,
  title={Energy-based backdoor defense against federated graph learning},
  author={Wan, Guancheng and Shi, Zitong and Huang, Wenke and Zhang, Guibin and Tao, Dacheng and Ye, Mang},
  booktitle={Proceedings of the International Conference on Learning Representations},
  year={2025}
}

@inproceedings{GNNGuard,
author = {Zhang, Xiang and Zitnik, Marinka},
booktitle = {Proceedings of the Advances in Neural Information Processing Systems},
pages = {9263--9275},
title = {GNNGuard: Defending Graph Neural Networks against Adversarial Attacks},
volume = {33},
year = {2020}
}

@inproceedings{Meta,
author = {Z{\"{u}}gner, Daniel and G{\"{u}}nnemann, Stephan},
booktitle = {Proceedings of the International Conference on Learning Representations},
title = {Adversarial Attacks on Graph Neural Networks via Meta Learning},
year = {2019}
}

@inproceedings{GARNET,
author = {Deng, Chenhui and Li, Xiuyu and Feng, Zhuo and Zhang, Zhiru},
booktitle = {Proceedings of the Machine Learning Research},
pages = {3:1----3:23},
title = {GARNET: Reduced-Rank Topology Learning for Robust and Scalable Graph Neural Networks},
volume = {198},
year = {2022}
}

@inproceedings{GAT,
author = {Petar, Velickovic and Guillem, Cucurull and Arantxa, Casanova and Adriana, Romero and Pietro, Li{\`{o}} and Yoshua, Bengio},
booktitle = {Proceedings of the International Conference on Learning Representations},
title = {Graph Attention Networks},
year = {2018}
}

@article{LLM4FGL,
  title={Data-centric federated graph learning with large language models},
  author={Yan, Bo and Zhang, Zhongjian and Sun, Huabin and Zhang, Mengmei and Cao, Yang and Shi, Chuan},
  journal={arXiv preprint arXiv:2503.19455},
  year={2025}
}

@article{LGDUMAP,
  title={LLM-Guided Dynamic-UMAP for Personalized Federated Graph Learning},
  author={Puppala, Sai and Hossain, Ismail and Alam, Md Jahangir and Ahad, Tanzim and Talukder, Sajedul},
  journal={arXiv preprint arXiv:2511.09438},
  year={2025}
}

@article{pFedLoRA,
  title={pFedLoRA: Model-heterogeneous personalized federated learning with LoRA tuning},
  author={Yi, Liping and Yu, Han and Wang, Gang and Liu, Xiaoguang and Li, Xiaoxiao},
  journal={arXiv preprint arXiv:2310.13283},
  year={2023}
}

@inproceedings{FedExLoRA,
  title={FedEx-LoRA: Exact aggregation for federated and efficient fine-tuning of large language models},
  author={Singhal, Raghav and Ponkshe, Kaustubh and Vepakomma, Praneeth},
  booktitle={Proceedings of the Annual Meeting of the Association for Computational Linguistics},
  pages={1316--1336},
  year={2025}
}

@inproceedings{FedAMoLE,
  title={Personalized federated fine-tuning for llms via data-driven heterogeneous model architectures},
  author={Zhang, Yicheng and Qin, Zhen and Wu, Zhaomin and Hou, Jian and Deng, Shuiguang},
  booktitle={Proceedings of the ACM Web Conference},
  pages={5099--5110},
  year={2026}
}

@inproceedings{FedALT,
  title={Fedalt: Federated fine-tuning through adaptive local training with rest-of-world lora},
  author={Bian, Jieming and Wang, Lei and Zhang, Letian and Xu, Jie},
  booktitle={Proceedings of the AAAI Conference on Artificial Intelligence},
  pages={19728--19736},
  year={2026}
}

@article{llama24,
  title={The llama 3 herd of models},
  author={Grattafiori, Aaron and Dubey, Abhimanyu and Jauhri, Abhinav and Pandey, Abhinav and Kadian, Abhishek and Al-Dahle, Ahmad and Letman, Aiesha and Mathur, Akhil and Schelten, Alan and Vaughan, Alex and others},
  journal={arXiv preprint arXiv:2407.21783},
  year={2024}
}

\appendix
\section{Appendix}

\subsection{Algorithm}
\label{Ap_Alg}

Here we summarize one communication round of FedLSG in algorithmic form.
\begin{algorithm}[!h]
    \caption{FedLSG: One-Round Training and Aggregation}
    \label{algorithm1}
    \renewcommand{\algorithmicrequire}{\textbf{Input:}}
    \renewcommand{\algorithmicensure}{\textbf{Output:}}
    \begin{algorithmic}[1]
        \REQUIRE Participating clients $\{G_i\}_{i=1}^{N_c}$, global backbone parameters $\theta^t$, global LoRA prior $(A^t,B^t)$, suspicious-edge budget $K$, client learning rate $\eta$
        \ENSURE Updated global parameters $\theta^{t+1}$ and updated global LoRA prior $(A^{t+1},B^{t+1})$
        \STATE Server broadcasts $\theta^t$ and $(A^t,B^t)$ to all clients.
        \FOR{each client $i \in \{1,\dots,N_c\}$ in parallel}
            \STATE Initialize local backbone $\theta_i^t \leftarrow \theta^t$ and local LoRA factors $(A_i^t,B_i^t) \leftarrow (A^t,B^t)$.
            \STATE Compute SRGAT attention and local node representations by Eqs. \ref{Eq_sij}--\ref{Eq_Hi}.
            \STATE Compute rule edge anomaly scores $r_{uv}^{\mathrm{rule}}$ for all $(u,v)\in E_i$.
            \STATE Select top-$K$ suspicious edges $\mathcal{E}_i^{\mathrm{sel}}$ according to $r_{uv}^{\mathrm{rule}}$.
            \STATE Run the client-side graph-attentive semantic student and LoRA compensator by Eqs. \ref{Eq_student_gat}--\ref{Eq_student_lora}.
            \FOR{each $(u,v)\in\mathcal{E}_i^{\mathrm{sel}}$}
                \STATE Build sampled neighborhood query set $\mathcal{U}_{uv}^{\mathrm{edge}}$ by Eqs. \ref{Eq_ax_sampling}--\ref{Eq_ax_edge_queries}.
                \STATE Compute split semantic edge score $r_{uv}^{\mathrm{llm}}$ by Eq. \ref{Eq_ax_rsplit}.
                \STATE Fuse rule and semantic risks to obtain $\tilde r_{uv}$ by Eq. \ref{Eq_rfuse}.
                \STATE Update edge attention $\tilde s_{uv}$ by Eq. \ref{Eq_sematt}.
            \ENDFOR
            \STATE Build the client-level structural summary and rule trust $\tau_{i,\mathrm{client}}^{\mathrm{rule}}$.
            \STATE Build the cohort-aware semantic input and compute $\tau_{i,\mathrm{client}}^{\mathrm{llm}}$ by Eq. \ref{Eq_ax_client_split}.
            \STATE Solve for adaptive fusion weights $a_1,a_2$ by Eq. \ref{Eq_ab}, and obtain fused trust $\tilde\tau_{i,\mathrm{client}}$ by Eq. \ref{Eq_clientfuse}.
            \STATE Receive detached teacher targets from the server and compute $\mathcal{L}_{\mathrm{align}}$, $\mathcal{L}_{\mathrm{LoRA}}$, and $\mathcal{L}_{\mathrm{distill}}$ by Eqs. \ref{Eq_align}--\ref{Eq_distill}.
            \STATE Update $(\theta_i^t,A_i^t,B_i^t)$ using the total objective in Eq. \ref{Eq_splitloss}.
            \STATE Compute trust-aware aggregation weight $w_i^t$ by Eq. \ref{Eq_aggw}.
            \STATE Upload $\theta_i^t$, $(A_i^t,B_i^t)$, and $w_i^t$ to the server.
        \ENDFOR
        \STATE Server normalizes $\{w_i^t\}_{i=1}^{N_c}$ to obtain $\{\bar w_i^t\}_{i=1}^{N_c}$ by Eq. \ref{Eq_weightedavg}.
        \STATE Aggregate backbone parameters: $\theta^{t+1} \leftarrow \sum_{i=1}^{N_c}\bar w_i^t \theta_i^t$.
        \STATE Aggregate low-rank parameters $(A^{t+1},B^{t+1})$ by Eq. \ref{Eq_loraagg}.
        \RETURN $\theta^{t+1}$, $(A^{t+1},B^{t+1})$
    \end{algorithmic}
\end{algorithm}

\subsection{Complexity decomposition}
\label{Ax_complexity}

We first state the main conclusion. The overall complexity of FedLSG is of the same order as a \emph{FedLoRA+GAT}-style design: the client side is still dominated by the GAT/SRGAT backbone, while only lightweight low-rank semantic adaptation is synchronized. Compared with traditional FedGNNs, the main extra burden is one centralized server-side semantic teacher together with a bounded client-side student/LoRA branch. This is still far cheaper than LLM4FGL and LG-DUMAP, which replicate the full LLM stack on all clients.

We now rewrite the complexity according to the actual pipeline in the main text and Appendix 2. Let $N_n$ be the number of nodes on a client, $E$ the number of edges, $D$ the feature dimension, $C$ the number of classes, $H$ the hidden size, $K_e=|\mathcal{E}_i^{\mathrm{sel}}|$ the number of selected suspicious edges on client $i$, $N_c$ the number of participating clients, $L$ the total semantic depth, $k$ the number of client-resident student blocks, $r$ the LoRA rank, and $|\Theta|$ the number of trainable backbone parameters. Since each selected edge $(u,v)$ is queried $n_{uv}$ times under the one-neighbor-per-query strategy, we further define the total local edge-query budget
\begin{equation}
N_q = \sum_{(u,v)\in \mathcal{E}_i^{\mathrm{sel}}} n_{uv}.
\label{Eq_nq}
\end{equation}
We use $\mathcal{C}_{\mathrm{GAT}}$ to denote the baseline cost of one local training pass of the original graph backbone.

\paragraph{1) Structural backbone.} FedLSG preserves the original FedGNN training loop, so the dominant local cost is still the SRGAT backbone. Relative to the base GAT, the additional structural operation is the similarity-aware edge scoring in Eq. \ref{Eq_sij}, which requires one pass over local edges and node features. The local classification loss adds the usual $O(N_nC)$ term. Hence the per-client structural cost is
\begin{equation}
\mathcal{C}_{\mathrm{struct}} =
\mathcal{C}_{\mathrm{GAT}} + O(ED) + O(N_nC).
\label{Eq_cstruct_new}
\end{equation}
Thus, before any semantic reasoning is invoked, FedLSG already has the same leading profile as a standard FedGNN with a similarity-aware GAT backbone.

\paragraph{2) rule suspicious-edge selection and client summary construction.} The rule branch computes $r_{uv}^{\mathrm{rule}}$ from local homophily, community crossing, overlap, clustering support, and anomaly statistics. These quantities are all obtained from local neighborhoods and edge statistics, so the full edge scan costs
\begin{equation}
\mathcal{C}_{\mathrm{rule}} = O(ED + E + N_n).
\label{Eq_cheu_new}
\end{equation}
Selecting the top-$K_e$ suspicious edges can be implemented by partial sorting or a heap, costing at most
\begin{equation}
\mathcal{C}_{\mathrm{topK}} = O(E\log K_e).
\label{Eq_ctopk}
\end{equation}
In addition, the client-level structural summary in Eq. \ref{Eq_textclient} traverses local graph statistics, semantic-risk summaries, and one copy of the parameter divergence term, which gives
\begin{equation}
\mathcal{C}_{\mathrm{summary}} = O(N_n + E + |\Theta|).
\label{Eq_csummary_new}
\end{equation}

\paragraph{3) Client-side graph-student and LoRA adaptation.} After suspicious edges are selected, each client runs only a \emph{lightweight} graph-attentive semantic student with $k$ blocks, rather than the full semantic stack. From Eqs. \ref{Eq_student_gat}--\ref{Eq_student_lora}, the student first computes one semantic embedding and then performs $k$ local attention passes on the private graph, followed by a rank-$r$ LoRA correction over node states. Therefore, the client semantic encoding cost is upper bounded by
\begin{equation}
\mathcal{C}_{\mathrm{stu}} =
O\!\left(k(EH + N_nH^2)\right) + O(N_nrH).
\label{Eq_cstu}
\end{equation}
This is the core reason why FedLSG stays close to FedLoRA+GAT: only a shallow student and low-rank correction run on each client.

\paragraph{4) Edge-level split semantic scoring.} For each selected suspicious edge, FedLSG builds a deterministic structural text and then issues $n_{uv}$ short neighborhood queries, as described in Eqs. \ref{Eq_ax_edge_queries}--\ref{Eq_ax_rsplit}. Because the prompt format is fixed and each query appends only one sampled neighbor plus the local edge state $z_{uv}^{\mathrm{stu}}$, the total local edge-level semantic scoring cost is linear in the query budget:
\begin{equation}
\mathcal{C}_{\mathrm{edge\text{-}sem}} = O(N_qH).
\label{Eq_cedgesem}
\end{equation}
The averaging step in Eq. \ref{Eq_ax_rsplit} adds only $O(N_q)$ and does not change the asymptotic order. Since $N_q=\sum n_{uv}$ is controlled by the prompt budget and only applies to the top-$K_e$ suspicious edges, this term is explicitly bounded.

\paragraph{5) Client-level split trust scoring.} For trust-aware aggregation, FedLSG additionally serializes one compact behavioral summary per client and evaluates the client-level split trust in Eq. \ref{Eq_ax_client_split}. The local serialization cost is linear in the number of summarized statistics, while the cohort concatenation scales linearly with the number of participating clients. Hence the semantic trust-calibration overhead is
\begin{equation}
\mathcal{C}_{\mathrm{client\text{-}sem}} = O(N_cH) + O(N_c).
\label{Eq_cclientsem}
\end{equation}
The adaptive fusion in Eqs. \ref{Eq_rfuse}, \ref{Eq_clientfuse}, and \ref{Eq_ab} uses only two evidence sources and at most two views, so it contributes only constant-time arithmetic per edge or client and is negligible compared with graph propagation and semantic encoding.

\paragraph{6) Server-side teacher and distillation.} The heavy semantic depth never appears on clients. Instead, the server keeps the remaining $L-k$ semantic blocks and provides detached teacher targets for distillation. Since the teacher is used only for semantic supervision and trust calibration rather than for client-side deployment, its per-round server cost is
\begin{equation}
\mathcal{C}_{\mathrm{teacher}} =
O\!\left((N_q+N_c)(L-k)H^2\right).
\label{Eq_cteacher}
\end{equation}
After teacher targets are obtained, the client-side alignment and distillation terms in Eqs. \ref{Eq_align}--\ref{Eq_distill} require only vector comparisons between student outputs and detached targets. Their extra optimization cost is
\begin{equation}
\mathcal{C}_{\mathrm{distill}} =
O(N_nH + K_eH + rH).
\label{Eq_cdistill}
\end{equation}
Therefore, the expensive semantic depth is paid once on the server, not $N_c$ times on the clients.

\paragraph{7) Trust-aware aggregation and communication.} FedLSG aggregates the backbone parameters exactly as in weighted FedAvg and additionally aggregates only the low-rank factors $(A_i,B_i)$. Therefore the server-side aggregation and communication cost remains linear:
\begin{equation}
\mathcal{C}_{\mathrm{agg}} =
O\!\left(N_c(|\Theta| + rH)\right).
\label{Eq_cagg_new}
\end{equation}
The $O(N_c|\Theta|)$ term is the standard backbone aggregation cost, and the additional $O(N_crH)$ term comes only from synchronizing the low-rank factors.

\paragraph{Comparison with traditional FedGNNs and full-client LLM methods.} If we denote the complexity of a conventional FedGNN client by
\begin{equation}
\mathcal{C}_{\mathrm{FedGNN}} =
\mathcal{C}_{\mathrm{GAT}} + O(ED) + O(N_nC),
\label{Eq_cfedgnn_new}
\end{equation}
then FedLSG mainly adds the bounded rule-selection, shallow-student, query-budgeted semantic scoring, and low-rank synchronization terms in Eqs. \ref{Eq_cheu_new}--\ref{Eq_cagg_new}, plus one centralized teacher term in Eq. \ref{Eq_cteacher}. By contrast, if the full semantic depth were deployed on every client as in LLM4FGL or LG-DUMAP, the semantic cost would scale as
\begin{equation}
\mathcal{C}_{\mathrm{full\text{-}client\text{-}LLM}} =
O\!\left(N_c(N_q+N_n)LH^2\right),
\label{Eq_cfullclient_new}
\end{equation}
ignoring additional prompt-engineering and in-context-example overhead. FedLSG avoids this $N_c$-fold replication by keeping only $k \ll L$ lightweight blocks on each client and centralizing the remaining $L-k$ blocks on the server.

\paragraph{Overall complexity.} Combining the above terms, one communication round of FedLSG is upper bounded by
\begin{equation}
\begin{aligned}
\mathcal{C}_{\mathrm{FedLSG}} ={}&
\mathcal{C}_{\mathrm{GAT}}
+ O(ED + N_nC)
+ O(ED + E + N_n)
+ O(E\log K_e) \\
&+ O(N_n + E + |\Theta|)
+ O\!\left(k(EH + N_nH^2)\right)
+ O(N_nrH)
+ O(N_qH) \\
&+ O(N_cH + N_c)
+ O\!\left((N_q+N_c)(L-k)H^2\right)
+ O(N_nH + K_eH + rH) \\
&+ O\!\left(N_c(|\Theta| + rH)\right).
\end{aligned}
\label{Eq_ctotal_new}
\end{equation}
Since $k$ and $r$ are small constants, $K_e \ll E$ by construction, and $N_q$ is explicitly controlled by the prompt budget, the client-side leading term remains the original graph backbone plus a shallow LoRA-enhanced student. In short, the complexity of FedLSG is essentially ``FedLoRA+GAT + one server-side LLM teacher,'' which is slightly heavier than traditional FedGNNs but still much lighter than LLM4FGL and LG-DUMAP.

\subsection{Implementation Details for Semantic Calibration and Trust Estimation}
\label{Ax_sem_prompt}

This appendix details how FedLSG converts structural evidence into bounded semantic signals. The design follows a simple progression. We first derive cheap structural anomaly scores to identify \emph{where} semantic reasoning is needed. We then run a lightweight graph-attentive student on each client to produce \emph{how} those suspicious structures should be interpreted. Finally, we serialize client behavior into compact cohort-level summaries so that trust-aware aggregation can compare clients without exposing raw private graphs. Throughout this appendix, the key principle is that all structure-to-text mappings are deterministic, while trainable semantic adaptation appears only through the client-side graph-student and its low-rank LoRA factors.

\paragraph{Step 1: rule edge scoring for candidate selection.} For each edge $(u,v)$ on client $i$, FedLSG first computes a lightweight structural anomaly score
\begin{equation}
r_{uv}^{\mathrm{rule}} = \lambda^{r}_1 \,r_{uv}^{\mathrm{sim}}
+ \lambda^{r}_2 \,r_{uv}^{\mathrm{cross}}
+ \lambda^{r}_3 \,r_{uv}^{\mathrm{hom}}
+ \lambda^{r}_4 \,r_{uv}^{\mathrm{comm}}
+ \lambda^{r}_5 \,r_{uv}^{\mathrm{nbr}},
\label{Eq_rrule}
\end{equation}
where
\begin{equation}
r_{uv}^{\mathrm{sim}} = 1-\max\!\bigl(0,\cos(H_u,H_v)\bigr), \qquad
r_{uv}^{\mathrm{cross}} = \mathbb{I}[g_u \neq g_v],
\label{Eq_rparts1}
\end{equation}
\begin{equation}
r_{uv}^{\mathrm{hom}} = 1-h_{uv}, \qquad
r_{uv}^{\mathrm{comm}} = \min\!\left(1,\frac{\max(c_u,c_v)}{4}\right), \qquad
r_{uv}^{\mathrm{nbr}} = 1-\min(\eta_{uv}+j_{uv},1).
\label{Eq_rparts2}
\end{equation}
Here $g_u$ is the structural community ID of node $u$, $h_{uv}$ is the mean same-class neighbor ratio around the two endpoints, $c_u$ is the number of distinct neighboring communities around node $u$, $\eta_{uv}$ is a local clustering-support ratio, and $j_{uv}$ is the Jaccard overlap between endpoint neighborhoods. The resulting $r_{uv}^{\mathrm{rule}}\in[0,1]$ is not yet a semantic judgment. It is only a cheap structural filter used to rank edges and select the candidate set $\mathcal{E}_i^{\mathrm{sel}}$ for further reasoning.

\paragraph{Step 2: Local statistics for rule client trust.} FedLSG also extracts a client-level structural summary that will later serve as the non-semantic part of trust estimation. For client $i$, we compute
\begin{equation}
H_i^{\mathrm{label}} = - \frac{\sum_c p_{ic}\log(p_{ic}+\epsilon)}{\log C}, \qquad
H_i^{\mathrm{att}} = - \frac{\sum_{e\in E_i} q_{ie}\log(q_{ie}+\epsilon)}{\log |E_i|},
\label{Eq_entropy}
\end{equation}
\begin{equation}
\bar{r}_i = \frac{1}{|E_i|}\sum_{(u,v)\in E_i} \tilde{r}_{uv}, \qquad
\psi_i = \frac{1}{|E_i|}\sum_{(u,v)\in E_i}\mathbb{I}[\tilde{r}_{uv}>0.6],
\label{Eq_meanrisk}
\end{equation}
\begin{equation}
\Delta_i = \frac{1}{|\Theta|}\sum_{\ell\in\Theta}\left(1-\left|\cos(\theta_i^\ell,\theta_g^\ell)\right|\right), \qquad
\xi_i = |2S_1-1|,
\label{Eq_div_sparse}
\end{equation}
where $p_{ic}$ is the local label histogram, $q_{ie}$ is the normalized attention mass, $\Theta$ is the set of trainable parameter tensors, and $S_1$ is the proportion of nearly unchanged client-global parameters. We additionally summarize the local graph by average degree $\bar d_i$, density $\delta_i$, homophily $\phi_i$, cross-community ratio $\chi_i$, and structural anomaly $\omega_i$, and define the motif anomaly $m_i=\min(1,0.5\bar{r}_i+0.5\psi_i)$.

These statistics are combined into a local anomaly score
\begin{equation}
A_i^{\mathrm{local}} =
\lambda^{l}_1 \Delta_i + \lambda^{l}_2 \bar{r}_i + \lambda^{l}_3 \psi_i + \lambda^{l}_4 (1-H_i^{\mathrm{label}})
+ \lambda^{l}_5 m_i + \lambda^{l}_6 \xi_i + \lambda^{l}_7 (1-\phi_i) + \lambda^{l}_8 \chi_i + \lambda^{l}_9 \omega_i,
\label{Eq_clientanom_local}
\end{equation}
which yields a local rule trust value
\begin{equation}
\tau_i^{\mathrm{rule}} = 1-\rho_1 A_i^{\mathrm{local}}.
\label{Eq_taurule}
\end{equation}
This score captures whether a client looks unusual in isolation. However, because benign non-IID clients may also deviate from the average, aggregation ultimately requires a cohort-normalized version. We therefore define
\begin{equation}
A_i^{\mathrm{cohort}} =
\lambda^{c}_1 [z(\Delta_i)]_+ + \lambda^{c}_2 [z(\bar{r}_i)]_+ + \lambda^{c}_3 [z(\psi_i)]_+ + \lambda^{c}_4 [z(1-H_i^{\mathrm{label}})]_+,
\label{Eq_clientanom_cohort}
\end{equation}
where $[x]_+=\max(0,x)$ and $z(\cdot)$ is a robust client-normalized $z$-score. The corresponding rule trust for aggregation is
\begin{equation}
\tau_{i,\mathrm{client}}^{\mathrm{rule}} = 1-\rho_2 A_i^{\mathrm{cohort}}.
\label{Eq_clientrule}
\end{equation}

\paragraph{Step 3: Client-side graph-student semantic encoding.} After selecting $\mathcal{E}_i^{\mathrm{sel}}$, the client runs a lightweight graph-attentive semantic student on its private graph. The student first computes
\begin{equation}
H_i^{(0)} = W_e X_i, \qquad
H_i^{(\ell+1)} = \mathrm{SRGAT}\!\left(H_i^{(\ell)}, E_i\right), \quad \ell=0,\dots,k-1,
\label{Eq_student_gat}
\end{equation}
where $W_e$ is the semantic embedding layer and only $k$ lightweight attention blocks are retained on the client. A low-rank compensator then refines the student representation:
\begin{equation}
\Delta_i^{\mathrm{LoRA}} = A_iB_i, \qquad
Z_i^{\mathrm{stu}} = H_i^{(k)} + \Delta_i^{\mathrm{LoRA}}H_i^{(k)}.
\label{Eq_student_lora}
\end{equation}
For each selected edge $(u,v)\in \mathcal{E}_i^{\mathrm{sel}}$, we form the local semantic edge state
\begin{equation}
z_{uv}^{\mathrm{stu}} = \frac{1}{2}\left(Z_{i,u}^{\mathrm{stu}} + Z_{i,v}^{\mathrm{stu}}\right).
\label{Eq_student_edge}
\end{equation}
This representation is the only trainable semantic signal injected into edge-level prompting on the client.

\paragraph{Step 4: Deterministic structural serialization for edge-level semantic scoring.} FedLSG does not feed raw subgraphs to the split semantic branch. Instead, it converts each suspicious edge into a short deterministic structural record
\begin{equation}
\mathcal{T}_{uv}^{\mathrm{edge}} =
\Phi_{\mathrm{edge}}\!\left(
u,v,d_u,d_v,h_{uv},\chi_{uv},\omega_{uv},c_{uv},j_{uv},\eta_{uv}
\right),
\label{Eq_textedge}
\end{equation}
where $\Phi_{\mathrm{edge}}(\cdot)$ denotes a deterministic structural record constructor that packages edge-centered graph statistics into a fixed tuple, $h_{uv}$ denotes local homophily, $\chi_{uv}$ indicates whether the edge crosses communities, $\omega_{uv}$ is the edge anomaly score, $c_{uv}$ is the number of common neighbors, $j_{uv}$ is the Jaccard overlap, and $\eta_{uv}$ is the local clustering-support ratio.
The purpose of this step is to bridge the gap between graph statistics and semantic scoring. Up to this point, FedLSG has identified suspicious edges through structural rules and has computed a local student representation $z_{uv}^{\mathrm{stu}}$, but these numerical quantities are still heterogeneous in scale and format. Before the split semantic branch can reason over them, they must be converted into a stable textual interface that is consistent across clients and rounds. We therefore use a deterministic serializer that maps the structural record $\mathcal{T}_{uv}^{\mathrm{edge}}$ into short security-oriented clauses.

To keep this mapping reproducible and robust to small numerical fluctuations, each scalar is first bucketized into a small number of qualitative levels:
\begin{equation}
\operatorname{lvl}(x;\alpha,\beta)=
\begin{cases}
\texttt{low}, & x \le \alpha,\\
\texttt{moderate}, & \alpha < x < \beta,\\
\texttt{high}, & x \ge \beta,
\end{cases}
\label{Eq_bucket}
\end{equation}
with $(\alpha,\beta)=(0.25,0.75)$ by default. This bucketization prevents the prompt from overfitting to insignificant numeric noise while preserving the coarse anomaly pattern that is most useful for semantic discrimination. The edge serializer then verbalizes the bucketized statistics into short clauses, for example:
\begin{equation}
\begin{aligned}
h_{uv} &\mapsto \texttt{``Only }100h_{uv}\%\texttt{ of nearby neighbors share the same class signal.''},\\
\chi_{uv}=1 &\mapsto \texttt{``This connection crosses community boundaries.''},\\
\chi_{uv}=0 &\mapsto \texttt{``This connection stays within the same community.''},\\
\eta_{uv}\ge 0.45 &\mapsto \texttt{``The target edge lies inside a tightly connected local cluster.''},\\
\eta_{uv}<0.45 &\mapsto \texttt{``The target edge is weakly supported by the local neighborhood.''},\\
\omega_{uv} &\mapsto \texttt{``Structural anomaly score = }\omega_{uv}\texttt{.''}.
\end{aligned}
\label{Eq_edge_rules}
\end{equation}
The neighborhood-overlap evidence is verbalized separately as
\begin{equation}
\begin{aligned}
(c_{uv},j_{uv}) \mapsto
\texttt{``Neighbor overlap remains limited with Jaccard similarity }j_{uv} \\
\texttt{and }c_{uv}\texttt{ common neighbors.''}.
\end{aligned}
\label{Eq_edge_overlap_rule}
\end{equation}
As a result, $\Psi_{\mathrm{edge}}(\mathcal{T}_{uv}^{\mathrm{edge}})$ becomes a fixed serializer from structural statistics to short natural-language evidence, where $\Psi_{\mathrm{edge}}(\cdot)$ denotes the deterministic map from a structured edge record to its textual description. This output is not the final semantic score yet. Rather, it serves as the textual anchor of the next step: each serialized center edge will be combined with one sampled 1-hop neighbor description and the learned student state $z_{uv}^{\mathrm{stu}}$ to form the bounded prompt set used by the split semantic scorer.

\paragraph{Step 5: Sampled neighborhood prompting and edge-level split semantic score.} For one suspicious edge, directly concatenating the full neighborhood would be noisy and token-inefficient. FedLSG therefore follows a one-neighbor-per-query design. Let
\begin{equation}
\mathcal{C}_{uv}^{1\text{-hop}}=\mathcal{N}(u)\cup\mathcal{N}(v)\setminus\{u,v\},
\label{Eq_ax_candidate}
\end{equation}
and sample
\begin{equation}
w_m \sim \operatorname{Sample}\!\left(\mathcal{C}_{uv}^{1\text{-hop}}\right),
\qquad m=1,\dots,n_{uv},
\label{Eq_ax_sampling}
\end{equation}
where $n_{uv}$ is the prompt budget of edge $(u,v)$. Here each sampled $w_m$ is a concrete 1-hop neighbor node used to instantiate a neighbor-side structural record $\mathcal{T}_{w_m}^{\mathrm{nbr}}$ and its textual description $\Psi_{\mathrm{nbr}}(\mathcal{T}_{w_m}^{\mathrm{nbr}})$. In other words, $w_m$ is not passed directly to the semantic scorer; it first determines which local neighbor evidence is extracted around the suspicious center edge. We then build the sampled query set
\begin{equation}
\mathcal{U}_{uv}^{\mathrm{edge}}=
\{\mathcal{U}_{uv,m}^{\mathrm{edge}}\}_{m=1}^{n_{uv}},
\qquad
\mathcal{U}_{uv,m}^{\mathrm{edge}}=
\Psi_{\mathrm{edge}}(\mathcal{T}_{uv}^{\mathrm{edge}})
\oplus
\Psi_{\mathrm{nbr}}(\mathcal{T}_{w_m}^{\mathrm{nbr}})
\oplus
z_{uv}^{\mathrm{stu}},
\label{Eq_ax_edge_queries}
\end{equation}
so each query contains the center-edge summary, the description induced by the sampled neighbor $w_m$, and the local graph-student state. The final split semantic score is
\begin{equation}
r_{uv}^{\mathrm{llm}} =
\mathcal{F}_{\mathrm{LLM}}^{\mathrm{edge}}
\!\left(\mathcal{S}^{\mathrm{det}}, \mathcal{U}_{uv}^{\mathrm{edge}}, \Delta_i^{\mathrm{LoRA}}\right)
=
\frac{1}{n_{uv}}
\sum_{m=1}^{n_{uv}}
\operatorname{score}\!\left(
\mathcal{S}^{\mathrm{det}}, \mathcal{U}_{uv,m}^{\mathrm{edge}}, \Delta_i^{\mathrm{LoRA}}
\right),
\label{Eq_ax_rsplit}
\end{equation}
where $\mathcal{S}^{\mathrm{det}}$ is the same shared deterministic system-level instruction defined in the main text, and $\mathcal{F}_{\mathrm{LLM}}^{\mathrm{edge}}(\cdot,\cdot,\cdot)$ is the local split-semantic scorer induced by the client-side student and LoRA compensation. Averaging across sampled neighbor contexts makes the score less sensitive to any single noisy neighbor.

\paragraph{Step 6: Client-level structural summary for semantic trust calibration.} Edge-level semantic scores are sufficient for refining suspicious links, but aggregation requires judging whether an entire client behaves abnormally relative to the cohort. We therefore serialize each client into a compact behavioral record
\begin{equation}
\mathcal{T}_i^{\mathrm{client}} = \Phi_{\mathrm{client}}\!\left(
\bar d_i,\delta_i,\phi_i,\chi_i,\omega_i,H_i^{\mathrm{label}},\bar{r}_i,\psi_i,\Delta_i,S_1,m_i
\right),
\label{Eq_textclient}
\end{equation}
and map it into a deterministic summary
\begin{equation}
\mathcal{S}_i^{\mathrm{client}}
=
\Psi_{\mathrm{client}}\!\left(\mathcal{T}_i^{\mathrm{client}}\right).
\label{Eq_client_summary}
\end{equation}
Here $\Phi_{\mathrm{client}}(\cdot)$ denotes the client-level analogue of $\Phi_{\mathrm{edge}}(\cdot)$, namely a deterministic constructor that packages round-wise client statistics into a fixed structured record, and $\Psi_{\mathrm{client}}(\cdot)$ denotes the corresponding deterministic text serializer. The same bucketization rule in Eq. \ref{Eq_bucket} is reused here, but the focus shifts from one edge to the full round-wise client behavior. Typical mappings include
\begin{equation}
\begin{aligned}
\delta_i &\mapsto \texttt{``The local graph contains }\operatorname{lvl}(\delta_i;0.2,0.6)\texttt{ly connected nodes.''},\\
\phi_i &\mapsto \texttt{``Homophily = }\phi_i\texttt{.''},\\
\chi_i &\mapsto \texttt{``Cross-community ratio = }\chi_i\texttt{.''},\\
\omega_i &\mapsto \texttt{``Structural anomaly score = }\omega_i\texttt{.''},\\
H_i^{\mathrm{label}} &\mapsto \texttt{``Label entropy = }H_i^{\mathrm{label}}\texttt{.''},\\
\Delta_i &\mapsto \texttt{``Parameter divergence = }\Delta_i\texttt{.''}.
\end{aligned}
\label{Eq_client_rules}
\end{equation}
We further append deterministic conclusion clauses when unusually risky patterns are present:
\begin{equation}
\begin{aligned}
\phi_i < 0.3 &\Rightarrow \texttt{``Only a small fraction of local connections stay within the same class signal.''},\\
\chi_i > 0.6 &\Rightarrow \texttt{``Connections frequently cross community boundaries.''},\\
\omega_i > 0.7 &\Rightarrow \texttt{``The graph exhibits unusual structural behavior compared with benign clients.''}.
\end{aligned}
\label{Eq_client_conditional_rules}
\end{equation}

\paragraph{Step 7: Cohort-level semantic trust for aggregation.} To compare clients within one round, we concatenate all client summaries into a cohort prompt:
\begin{equation}
\mathcal{P}^{\mathrm{cohort}} =
\text{ClientSummaryInstruction}
\oplus
\bigcup_{i=1}^{N_c}\text{ClientCohortLine}_i,
\label{Eq_promptcohort}
\end{equation}
\begin{equation}
\text{ClientCohortLine}_i =
\Psi_{\mathrm{client}}\!\left(
\bar d_i,\delta_i,\phi_i,\chi_i,\omega_i,H_i^{\mathrm{label}},\Delta_i,\tau_{i,\mathrm{client}}^{\mathrm{rule}}
\right),
\label{Eq_clientline}
\end{equation}
\begin{equation}
\bigcup_{i=1}^{N_c}\text{ClientCohortLine}_i
=
\Psi_{\mathrm{cohort}}\!\left(\{\text{summary}_i\}_{i=1}^{N_c}\right),
\label{Eq_cohort_prompt_map}
\end{equation}
where $\bigcup$ denotes string concatenation rather than numerical summation. We also summarize the client-side semantic response by the mean split semantic score on selected suspicious edges:
\begin{equation}
\mathcal{U}^{\mathrm{client}} = \mathcal{P}^{\mathrm{cohort}} \oplus \nu_i^{\mathrm{LoRA}},
\label{Eq_Uclient}
\end{equation}
\begin{equation}
\nu_i^{\mathrm{LoRA}} =
\frac{1}{|\mathcal{E}_i^{\mathrm{sel}}|}
\sum_{(u,v)\in \mathcal{E}_i^{\mathrm{sel}}}
r_{uv}^{\mathrm{llm}}.
\label{Eq_userclient_cohort}
\end{equation}

Finally, the client-level split semantic trust is
\begin{equation}
\tau_{i,\mathrm{client}}^{\mathrm{llm}} =
\mathcal{F}_{\mathrm{llm}}^{\mathrm{client}}
\!\left(\texttt{``client-normalized client behavior summary.''}, \mathcal{U}^{\mathrm{client}}, i\right),
\label{Eq_ax_client_split}
\end{equation}
where $\mathcal{F}_{\mathrm{llm}}^{\mathrm{client}}(\cdot,\cdot,i)$ scores client $i$ relative to the cohort rather than in isolation. This distinction is important: a client may look unusual locally yet still be consistent with benign non-IID heterogeneity, which is exactly why semantic trust is computed at the cohort level before aggregation.

\paragraph{Algorithmic view of deterministic serialization.} Algorithm \ref{Alg_s2t} summarizes the deterministic structure-to-text conversion used above. Its role is not to replace the graph-student, but to ensure that the semantic branch receives compact, reproducible, security-oriented descriptions instead of raw graph objects.

\begin{algorithm}[t]
\caption{Structure-to-Text Serializer}
\label{Alg_s2t}
\begin{algorithmic}[1]
\REQUIRE Structural record $\mathcal{R}$, stage type $z \in \{\texttt{edge}, \texttt{local-client}, \texttt{cohort}\}$, rule library $\Psi_z$
\ENSURE Natural-language structural summary $\mathcal{S}_z$
\STATE Initialize an empty sentence list $\mathcal{L} \leftarrow [\,]$
\FOR{each scalar field $x$ in $\mathcal{R}$}
    \STATE Bucketize $x$ by Eq. \ref{Eq_bucket} when $x$ is not emitted as an exact percentage or score
\ENDFOR
\IF{$z=\texttt{edge}$}
    \STATE Append cluster-support sentence from $\eta_{uv}$
    \STATE Append homophily sentence from $h_{uv}$
    \STATE Append community-boundary sentence from $\chi_{uv}$
    \STATE Append overlap sentence from $(c_{uv}, j_{uv})$
    \STATE Append anomaly sentence from $\omega_{uv}$
\ELSIF{$z=\texttt{local-client}$}
    \STATE Append density/connectivity sentence from $\delta_i$
    \STATE Append scalar-report sentences for $(\bar d_i,\phi_i,\chi_i,\omega_i,H_i^{\mathrm{label}},\Delta_i)$
    \STATE Append conclusion clauses triggered by Eq. \ref{Eq_client_conditional_rules}
\ELSIF{$z=\texttt{cohort}$}
    \FOR{each client record $\mathcal{R}_i$ in the cohort}
        \STATE Serialize $\mathcal{R}_i$ with $\Psi_{\mathrm{client}}$
        \STATE Append the rule cohort-trust line from $\tau_{i,\mathrm{client}}^{\mathrm{rule}}$
    \ENDFOR
\ENDIF
\STATE Concatenate all sentences in $\mathcal{L}$ with fixed ordering and line breaks
\STATE \textbf{return} $\mathcal{S}_z$
\end{algorithmic}
\end{algorithm}

Algorithm \ref{Alg_s2t} clarifies the logic of Appendix 2. First, structure is compressed into deterministic textual evidence. Second, the graph-student contributes only compact local semantic states and low-rank compensation. Third, the final semantic trust signal is produced only after the client is positioned relative to its cohort. This design makes the semantic branch interpretable, bounded, and compatible with privacy-preserving federated training.

\subsection{Derivation of the Adjustment Mechanism}
\label{Ax_adj}

We solve Eq. \ref{Eq_minf_} by applying the augmented lagrangian function: 
\begin{equation} \label{Eq_f_app}
\begin{aligned}
f = &\sum_{q=1}^{Q} \left| \left( \sum_{p=1}^{P}  a_p M_p^q(rule, llm) \right) - M_p^q(rule, llm) \right|
    + \frac{\lambda}{P} \left( \sum_{p=1}^{P} a_p^2 - \gamma \right).  
\end{aligned}
\end{equation}

Then, we take the derivative of Eq. \ref{Eq_f_app}, yielding:
\begin{equation}
\left\{\begin{aligned}
  & \frac{\partial f}{\partial a_p} = \sum_{q=1}^{Q} (\pm) M_p^q(rule, llm) + \frac{2\lambda}{P} a_p = 0, \\
  & \frac{\partial f}{\partial \lambda} = \frac{1}{P} \sum_{p=1}^{P} a_p^2 - \gamma = 0. \\
\end{aligned} \right.
\end{equation}


\begin{equation}
\left\{ \begin{aligned}
  & \frac{(\pm) \sum_{q=1}^{Q} M_1^q(rule, llm)}{-a_1} = \frac{2\lambda}{P}, \\
  & \frac{(\pm) \sum_{q=1}^{Q} M_2^q(rule, llm)}{-a_2} = \frac{2\lambda}{P}, \\
  & a_1^2 + a_2^2 - \gamma P = 0. \\
\end{aligned} \right.
\end{equation}

\begin{equation}
\left\{ \begin{matrix}
   \dfrac{\pm \sum\limits_{q=1}^{Q} M_1^q(rule, llm)}{-a_1} = \dfrac{\pm \sum\limits_{q=1}^{Q} M_2^q(rule, llm)}{-a_2}, \\
   a_1^2 + a_2^2 - \gamma P = 0. \\
\end{matrix} \right.
\end{equation}

Squaring both sides of the first term in the above equation gives:

\begin{equation}
\left\{ \begin{matrix}
   \dfrac{\left( \sum\limits_{q=1}^{Q} M_1^q(rule, llm) \right)^2}{a_1^2} = \dfrac{\left( \sum\limits_{q=1}^{Q} M_2^q(rule, llm) \right)^2}{a_2^2}, \\
   a_2^2 = \gamma P - a_1^2. \\
\end{matrix} \right.
\end{equation}

By combining the two terms in the above equation and obtain:
\begin{equation}
\frac{\left( \sum\limits_{q=1}^{Q} M_1^q(rule, llm) \right)^2}{a_1^2}
= \frac{\left( \sum\limits_{q=1}^{Q} M_2^q(rule, llm) \right)^2}{\gamma P - a_1^2}.
\end{equation}

\begin{equation}
(\gamma P - a_1^2) \left( \sum_{q=1}^{Q} M_1^q(rule, llm) \right)^2 = a_1^2 \left( \sum_{q=1}^{Q} M_2^q(rule, llm) \right)^2.
\end{equation}


\begin{equation}
\alpha = a_1 = \frac{\sqrt{\gamma P} \cdot \sum_{q=1}^Q M_1^q(rule, llm)}{\sqrt{\sum_{p=1}^P \left( \sum_{q=1}^Q M_p^q(rule, llm) \right)^2}}.
\end{equation}

In the same way, we have:

\begin{equation}
\beta = a_2 = \frac{\sqrt{\gamma P} \cdot \sum_{q=1}^Q M_2^q(rule, llm)}{\sqrt{\sum_{p=1}^P \left( \sum_{q=1}^Q M_p^q(rule, llm) \right)^2}}.
\end{equation}

Prove completion.

\subsection{Baselines}

\begin{enumerate}
\item SBA demonstrates that poisoning training graphs with predefined subgraph triggers can effectively implant backdoors into GNN models. 
\item UGBA introduces an adaptive trigger generation strategy that produces triggers closely resembling target nodes, thereby improving attack effectiveness while reducing detectability. 
\item DPGBA enhances stealthiness by generating triggers that preserve the original graph distribution, ensuring that the injected patterns remain statistically consistent with the underlying data. In FL scenarios, the attack surface becomes even broader. 
\item Some works explore both centralized and distributed backdoor attack strategies that exploit the aggregation mechanism in federated GNNs. They establish a comprehensive benchmark that systematically evaluates backdoor attacks in federated graph learning, facilitating more standardized comparisons across methods.
\item Prune and its variant PruneLD eliminate edges between dissimilar nodes based on feature inconsistency, aiming to filter out potential trigger connections. 
\item OD adopts a different perspective by leveraging autoencoder reconstruction errors to detect anomalous patterns introduced by backdoor attacks.
\item NoisyGCN injects controlled noise into the graph structure to expose unstable or poisoned edges, thereby improving detection capability.
\item RIGBD integrates randomized edge dropping with robust training strategies to mitigate the impact of backdoor triggers. 
\item FedTGE adopts an energy-based formulation to reweight client updates, assigning lower importance to anomalous contributions.
\end{enumerate}

\subsection{Parameter Settings}

We summarize the parameter settings used for Citeseer, Pubmed, Flickr, and Ogb-arxiv according to the released implementation. Unless otherwise stated, all datasets share the same optimization backbone: Adam is used as the optimizer, the local learning rate is set to $0.005$, and the SRGAT backbone uses two attention-bearing layers. The hidden dimension is selected from $\{128,256\}$ according to the dataset scale. For the \textbf{server-side full LLM}, we use the same bounded semantic configuration on all four datasets so that semantic guidance remains auxiliary rather than dominant: the maximum number of refined suspicious edges is $256$, the number of exposed salient feature indices is $6$, the trust floor is $0.1$, the heuristic--semantic mixing weight is $0.1$, the constraint term is fixed to $\gamma=1$, the global semantic effect scale is $0.05$, the edge-level semantic influence is $0.03$, the local trade-off reallocation strength is $0.03$, the aggregation-level trust influence is $0.08$, and the warm-up period is $2$ rounds. For the \textbf{client-side LoRA branch}, the hidden dimension is $128$, the total semantic depth is $4$, the number of client-resident semantic blocks is $2$, the LoRA rank is $8$, the semantic dropout is $0.1$, the trust-fusion hidden dimension is $128$, the alignment weight is $0.1$, the LoRA regularization weight is $10^{-4}$, the distillation weight is $0.05$, and the peer-trust consistency weight is $0.25$.

For \textbf{Citeseer}, we use a relatively lightweight configuration because the graph is small and sparse. The hidden dimension is set to $128$, the dropout rate is set to $0.1$, and the weight decay is set to $5\times 10^{-4}$. In the released training notes, Citeseer is treated as the default small-scale setting, so this dataset follows the standard optimization configuration without extra regularization or scaling adjustments. For \textbf{Pubmed}, the backbone width is also fixed to $128$, while the number of attention heads is set to $4$. The weight decay remains $5\times 10^{-4}$, but the dropout rate is adjusted according to the attack scenario: it is set to $0.2$ for SBA, $0.3$ for DPGBA, and $0.1$ or $0.5$ for UGBA. For \textbf{Flickr}, which is substantially larger and denser than Citeseer and Pubmed, we keep the hidden dimension at $128$ and use $4$ attention heads, but remove explicit regularization by setting the dropout rate to $0$ and the weight decay to $0$. For \textbf{Ogb-arxiv}, we use the largest backbone among the four datasets: the hidden dimension is increased to $256$, the number of attention heads is reduced to $1$, the dropout rate is set to $0.3$, the weight decay is set to $0$, and the number of local training epochs is increased to $200$. This configuration is explicitly marked in the implementation notes as the one that gives the strongest clean accuracy on Ogb-arxiv, making it the default large-scale setting for our reported experiments.

\subsection{Impact of LLM Selection}

Since FedLSG relies on LLM-based semantic reasoning to provide edge-level and client-level trust estimation, the choice of LLM may influence the final defense performance. To investigate this effect, we evaluate four representative instruction-tuned LLMs with different model sizes and architectures, including Qwen2.5-0.5B-Instruct, Llama-3.1-8B-Instruct, Gemma-3-12B-IT, and Qwen-3-14B. All models are integrated into FedLSG as the semantic teacher while keeping other experimental settings unchanged. We evaluate these LLMs under the DPGBA attack on two representative datasets, Pubmed and Ogb-arxiv. The attack success rate (ASR) and clean classification accuracy (ACC) are reported in Table~\ref{Tab_LLM_selection}. A lower ASR indicates stronger backdoor resistance, while a higher ACC reflects better preservation of benign graph information.

The results show that different LLMs achieve comparable performance in reducing ASR, indicating that FedLSG is relatively insensitive to the specific choice of semantic model for backdoor suppression. Even the lightweight Qwen2.5-0.5B-Instruct achieves competitive defense performance, demonstrating that semantic guidance rather than model scale itself plays the dominant role in identifying suspicious graph patterns. However, larger LLMs consistently provide improvements in clean accuracy. Specifically, Qwen-3-14B and Gemma-3-12B-IT achieve higher ACC compared with smaller models on both datasets. This suggests that larger models possess stronger semantic understanding capabilities, enabling more accurate preservation of benign structural information while suppressing malicious patterns. Therefore, increasing LLM capacity mainly benefits the accuracy-defense trade-off rather than further reducing ASR. Considering both effectiveness and computational cost, FedLSG can flexibly adopt different LLM backbones according to deployment requirements.







\begin{table}[h]
\centering
\caption{Impact of different LLMs under DPGBA attack.}
\label{Tab_LLM_selection}
\footnotesize
\begin{tabular}{lcccc}
\toprule
\multirow{2}{*}{LLM} 
& \multicolumn{2}{c}{Pubmed}
& \multicolumn{2}{c}{Ogb-arxiv}\\
\cline{2-5}
& ASR$\downarrow$ & ACC$\uparrow$
& ASR$\downarrow$ & ACC$\uparrow$\\
\midrule

Qwen2.5-0.5B-Instruct
&8.24$\pm$0.71&85.92$\pm$0.46
&32.85$\pm$0.82&65.88$\pm$0.67\\

Llama-3.1-8B-Instruct
&7.61$\pm$0.66&86.14$\pm$0.31
&30.21$\pm$0.68&66.22$\pm$0.78\\

Gemma-3-12B-IT
&7.53$\pm$0.65&87.37$\pm$0.42
&30.94$\pm$0.76&66.91$\pm$0.59\\

Qwen-3-14B
&7.25$\pm$0.58&88.42$\pm$0.29
&29.87$\pm$0.63&67.47$\pm$0.72\\

\bottomrule
\end{tabular}
\end{table}

\subsection{Ablation study}


\begin{figure}[h]
\centering
\includegraphics[width=0.60\columnwidth]{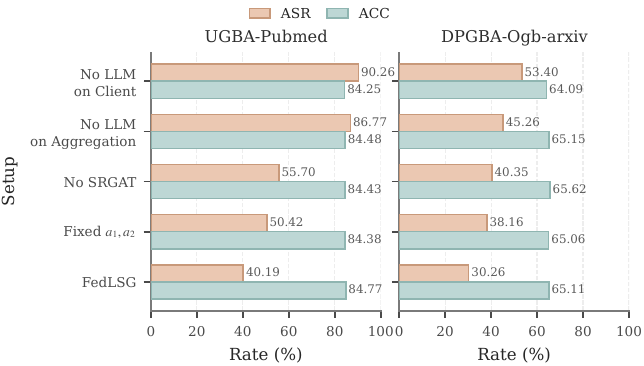}
\caption{Ablation study}
\label{Fig_Abl}
\end{figure}

To further investigate the contribution of each component in FedLSG, we conduct ablation experiments on Pubmed under UGBA and on Ogb-arxiv under DPGBA. The results are reported in Figure \ref{Fig_Abl}. The ablation settings are designed to evaluate the effectiveness of the graph backbone, the semantic reasoning branch, and the adaptive fusion mechanism.

Specifically, four variants are considered: (1) \emph{No LLM on Client}, where the client-side trust estimation is removed, and only the SRGAT backbone with LoRA-based adaptation is retained; (2) \emph{No LLM on Aggregation}, where the server-side trust estimation is removed during aggregation, and client updates are aggregated using only structural information; (3) \emph{No SRGAT}, where the similarity-aware attention mechanism is replaced with a standard GAT, removing edge-level similarity modeling; and (4) \emph{Fixed $a_1,a_2$}, where the adaptive weighting mechanism for rule and semantic evidence is disabled and replaced with fixed fusion coefficients ($a_1=a_2=0.5$).

The results in Figure \ref{Fig_Abl} lead to the following observations. First, removing the client-side trust estimation leads to a substantial increase in ASR, indicating that local graph-attentive semantic reasoning plays a critical role in identifying suspicious structures that cannot be captured by purely rule signals. Second, removing the server-side trust estimation also degrades performance, demonstrating that semantic guidance is essential for distinguishing benign non-IID heterogeneity from malicious updates during global model fusion. Third, replacing SRGAT with a standard GAT causes a clear performance drop under both attacks. This confirms that similarity-aware attention is effective in suppressing potential trigger edges and provides a strong structural foundation for downstream semantic calibration. Finally, using fixed fusion weights significantly weakens defense performance compared with the adaptive strategy, showing that static combination of rule and semantic evidence cannot properly handle varying attack strengths and distribution shifts across clients.

Overall, the full FedLSG model consistently achieves the best trade-off between attack mitigation and clean accuracy, validating the necessity of jointly modeling similarity-aware structure learning, graph-attentive semantic reasoning, and adaptive trust calibration.

\subsection{Hyperparameter Sensitivity Analysis}

\begin{figure}[h]
\centering
\includegraphics[width=0.99\columnwidth]{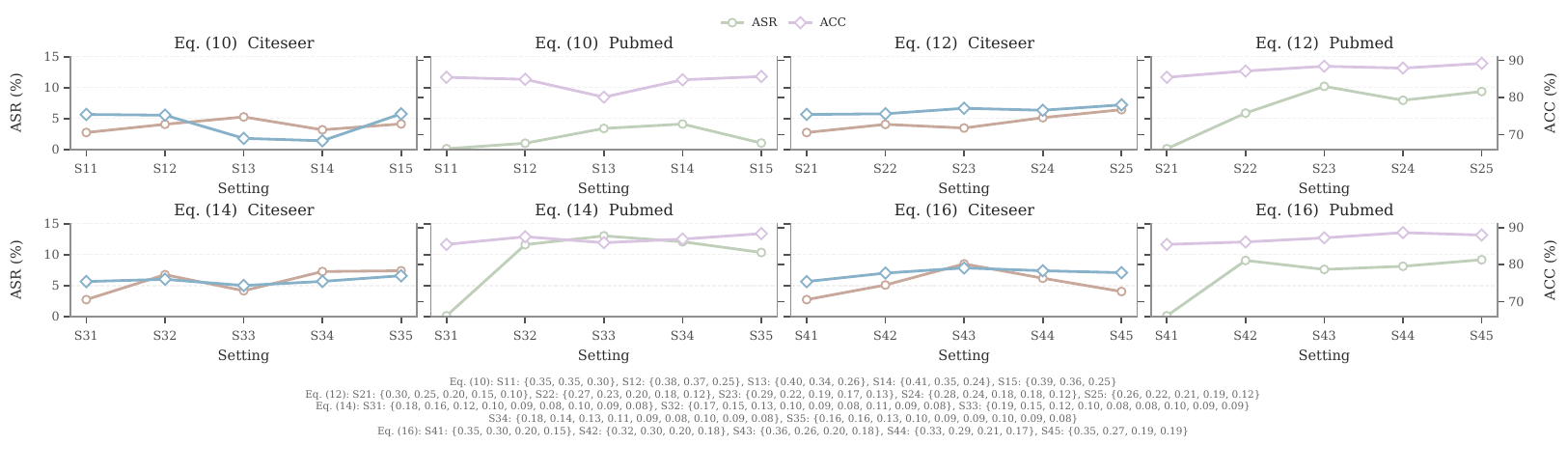}
\caption{Hyperparameter Sensitivity Analysis}
\label{Fig_Para}
\end{figure}

This subsection evaluates the sensitivity of the main coefficients in Eq. \ref{Eq_splitloss}, Eq. \ref{Eq_rrule}, Eq. \ref{Eq_clientanom_local}, and Eq. \ref{Eq_clientanom_cohort} under UGBA on Citeseer and SBA on Pubmed. Multiple coefficient settings are tested for each equation, while all other experimental configurations remain unchanged. The goal is to examine the robustness of the proposed method to coefficient variations and identify the best-performing configuration.

As shown in Figure \ref{Fig_Para}, the setting S11, S21, S31, and S41 consistently achieve the best overall trade-off between ASR and ACC, and are therefore selected as the default hyperparameters. Deviations from these configurations generally lead to higher ASR together with slightly higher ACC, which is mainly because the model becomes less regularized under imbalanced coefficient allocations. In such cases, certain components dominate the optimization process, improving task-specific fitting and thus marginally boosting ACC, but at the same time weakening the suppression of backdoor-related signals, which results in increased ASR. Overall, the results demonstrate that the model maintains strong robustness under both UGBA and SBA attacks, and the selected SX1 configurations provide a reliable and effective balance between attack resistance and classification performance.

\subsection{Impact of Client Scale}

\begin{figure}[h]
\centering
\includegraphics[width=0.89\columnwidth]{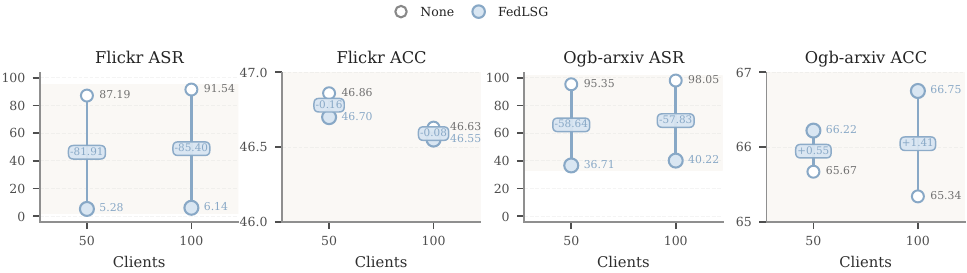}
\caption{Impact of Client Scale on Backdoor Defense Effectiveness}
\label{Fig_Client}
\end{figure}

We further examine the impact of scaling up the number of clients. Specifically, the client population is increased from 20 to 50 and 100. Experiments are conducted on Flickr and Ogb-arxiv under DPGBA attack, while keeping the proportion of malicious clients fixed at 60\%. The results in Figure \ref{Fig_Client} show that ASR increases as the client population grows for both FedLSG and the no-defense baseline. This phenomenon can be attributed to the fact that larger-scale participation leads to more diverse local optimization trajectories, which makes poisoned updates less distinguishable from benign ones in the aggregation space. As a result, malicious updates are more likely to bypass simple aggregation-level suppression mechanisms.

Despite this challenge, FedLSG consistently achieves significantly lower ASR across all client scales. This robustness stems from its dual-level defense design: the server-side semantic teacher identifies abnormal client behavior through semantic consistency analysis, while the client-side LoRA-based semantic students provide edge-level risk feedback to suppress malicious propagation during message passing. These two complementary mechanisms jointly reduce the influence of poisoned clients even as the system scales.

\subsection{Defense on heterogeneous graphs}
\label{Ax_heter}

\begin{figure*}[h]
\captionsetup[subfigure]{skip=10pt} 
    \centering
    \begin{subfigure}[b]{0.32\textwidth}
        \includegraphics[width=\linewidth]{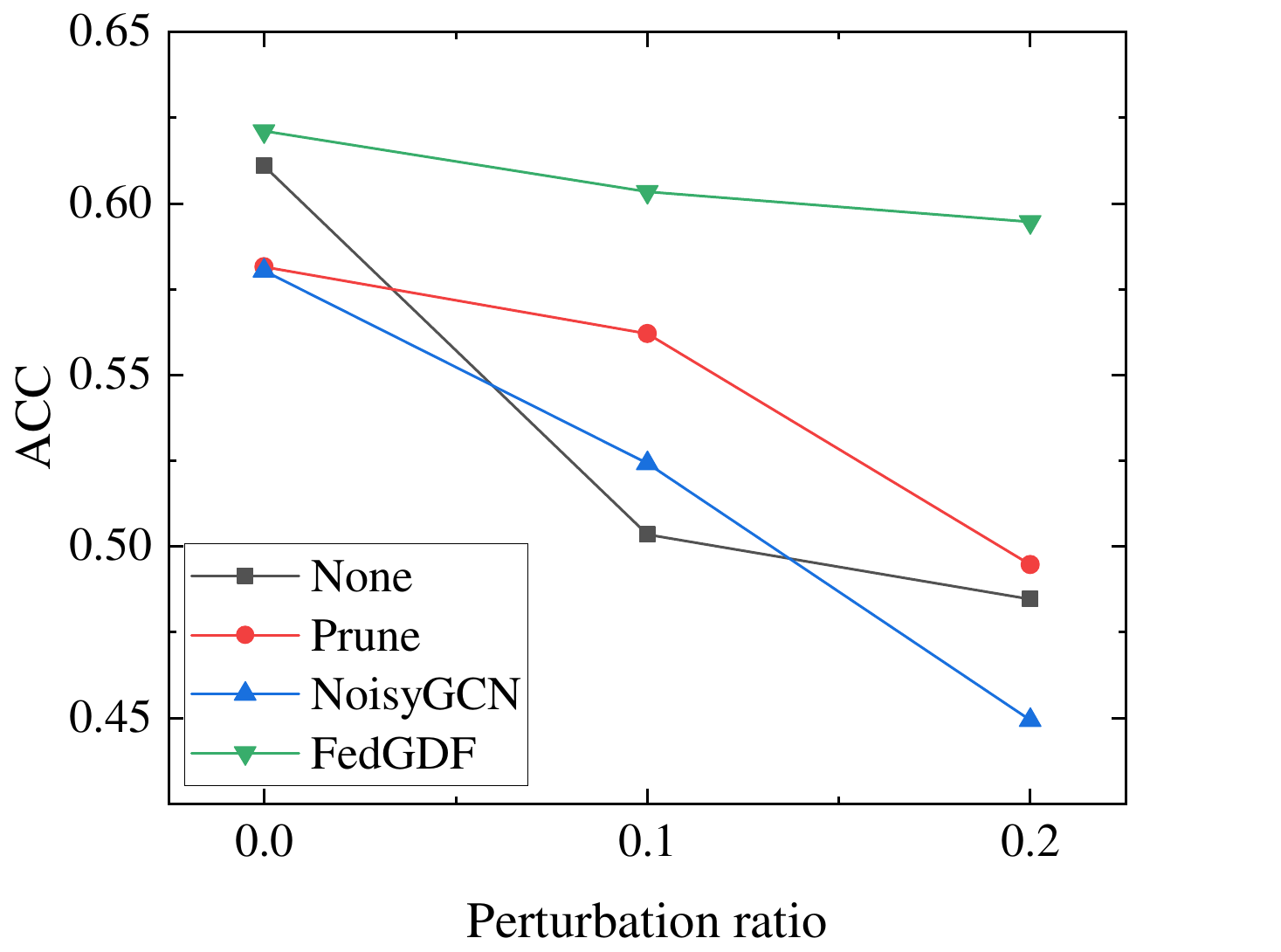}
        \caption{Meta attack}
    \end{subfigure}
    \hspace{0em}
    \begin{subfigure}[b]{0.31\textwidth}
        \includegraphics[width=\linewidth]{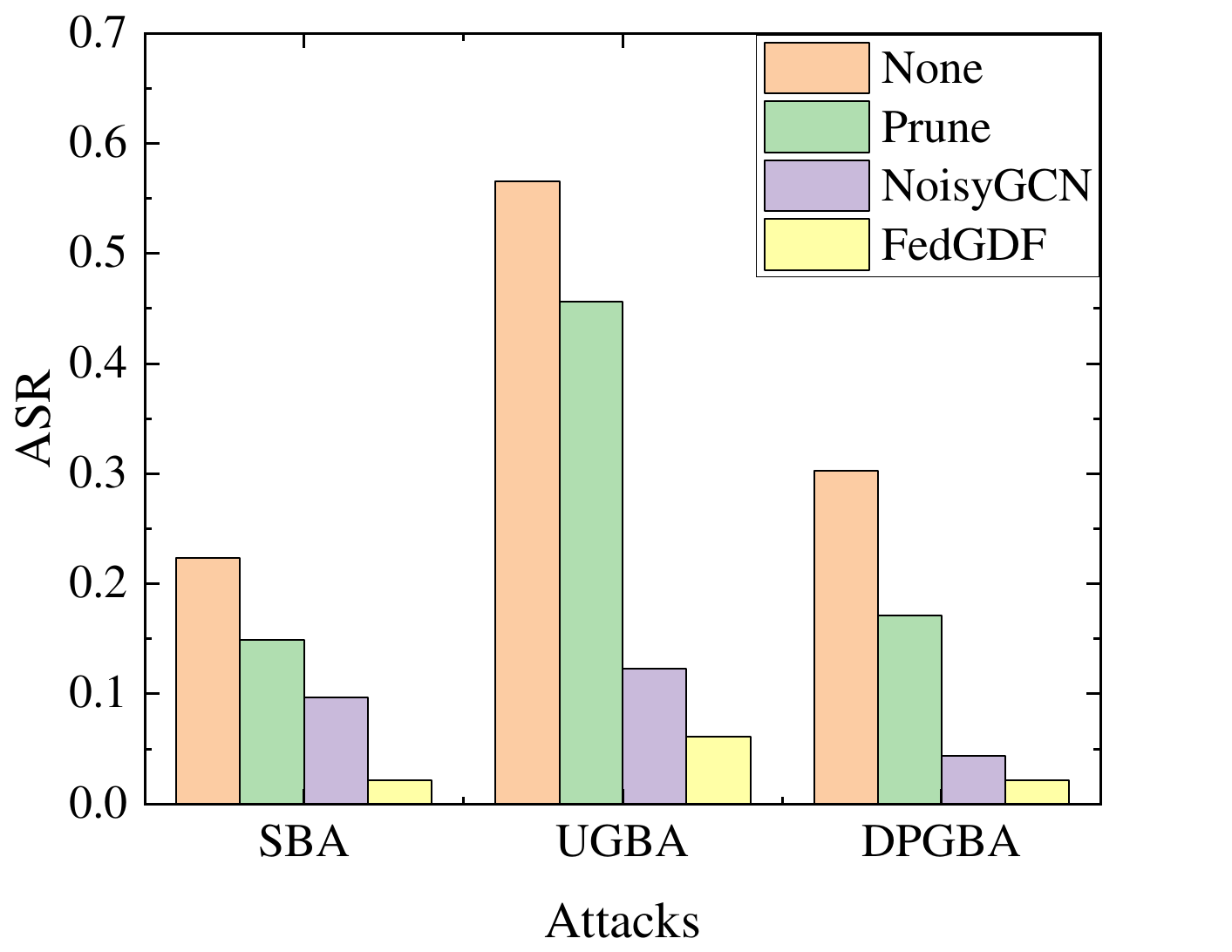}
        \caption{Backdoor attacks (ASR)}
    \end{subfigure}
    \hspace{0em}
    \begin{subfigure}[b]{0.32\textwidth}
        \includegraphics[width=\linewidth]{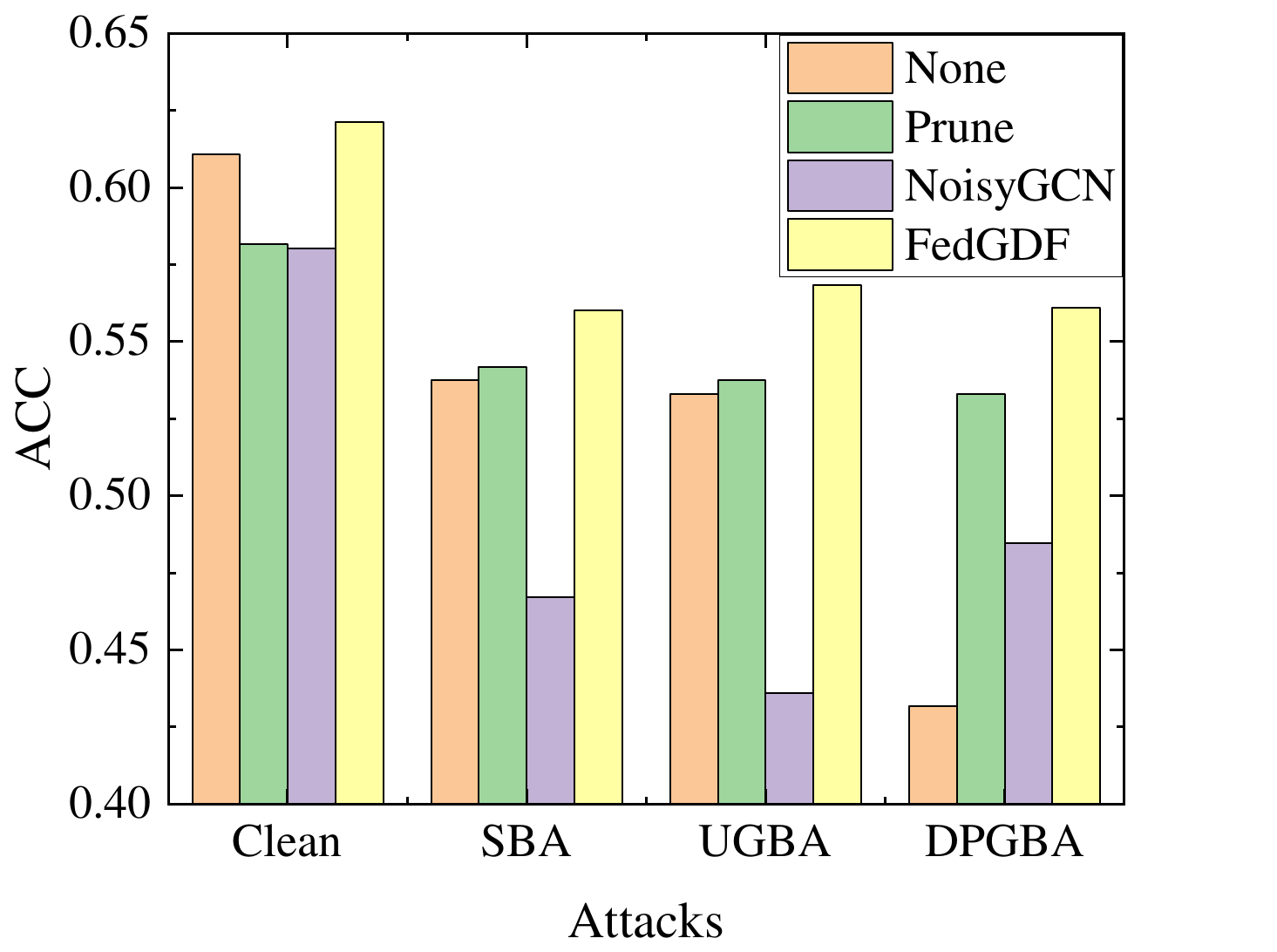}
        \caption{Backdoor attacks (ACC)}
    \end{subfigure}
    \caption{Defense on heterogeneous graphs}
    \label{Fig_Hetero}
\end{figure*}

We evaluate FedLSG on the heterogeneous graph dataset Chameleon \cite{GARNET}. Chameleon contains 2277 nodes, 31396 edges, 2325 feature dimensions, 5 classes. The dataset is split into 1,365 training, 455 validation, and 457 test samples. We attack Chameleon using backdoor attacks and the untargeted poisoning attack Meta \cite{Meta}. Meta directly reduces test accuracy by modifying the graph structure via meta-gradients. And we use the method from \cite{GNNGuard} to generate graphlet degree vectors. The results in Figure \ref{Fig_Hetero} show that backdoor attacks are not applicable to heterogeneous graphs. Since these attacks rely on the homogeneous graph assumption leads to considerable ASR, but at the expense of clean accuracy. This makes attacks easier to detect. For defending Meta, it can be seen that as the perturbation ratio increases, the baselines' ACC is significantly reduced, since they also rely on the homogeneous graph assumption. While FedLSG overcomes this limitation and nearly recovers the original performance using graphlet degree vectors.

\subsection{Non-IID-Louvain and IID settings}

All the above experiments are conducted under the non-identically distributed Louvain (Non-IID-Louvain) data partition setting. To comprehensively evaluate the robustness of FedLSG, we conduct a comparison between independent and identically distributed (IID) and Non-IID-Louvain settings in this section to comprehensively assess the FedLSG's defenses. We conduct experiments on all four datasets. The experimental results are presented in Table \ref{Tab_IID}, from which we draw the following conclusions: 

The models trained under IID settings are consistently more vulnerable to backdoor attacks compared to those trained under Non-IID-Louvain settings. This indicates that the IID setting actually constitutes a more challenging scenario for backdoor defense. The reason is that similar data distributions across clients in IID settings make clean clients more susceptible to being influenced by poisoned updates, thereby amplifying the attack effect. In contrast, the diversity introduced by non-IID partitions can partially mitigate such interference, leading to improved robustness, although often at the cost of reduced clean accuracy.

This observation is consistent with findings in \cite{Bkdfedgnn} and \cite{FedTGE}. Importantly, it also highlights the effectiveness of FedLSG across different data distributions. In particular, although IID settings are more vulnerable to backdoor attacks, FedLSG consistently maintains strong defense performance in both IID and Non-IID-Louvain scenarios. This indicates that FedLSG does not rely on distributional heterogeneity for robustness, but instead benefits from its semantic-aware modeling and client behavior characterization, which remain effective even when client data distributions are highly homogeneous.

\begin{table}[h]
\centering
\caption{Non-IID-Louvain and IID experiment results}
\label{Tab_IID}
\begin{tabular}{clcccc}
\toprule
Attack-       & \multirow{2}{*}{Defense} & \multicolumn{2}{c}{Non-IID} & \multicolumn{2}{c}{IID} \\ \cline{3-6}
Dataset    &    & ASR$\downarrow$     & ACC$\uparrow$     & ASR$\downarrow$           & ACC$\uparrow$          \\  \hline
SBA-          & None                     & 47.82       & 73.48      & 48.57             & 73.69            \\
Citeseer   & FedLSG                   & 0.00       & 74.97      & 0.00             & 74.54            \\
UGBA-         & None                     & 95.23      & 84.19      & 99.78            & 85.20            \\
Pubmed     & FedLSG                   & 42.30      & 85.13      & 8.79             & 85.85            \\
UGBA-         & None                     & 83.26      & 45.10      & 90.04            & 46.25            \\
Flickr     & FedLSG                   & 0.00       & 46.29      & 0.00             & 46.70            \\
DPGBA-        & None                     & 95.35      & 65.67      & 99.86            & 65.89            \\
Ogb-arxiv & FedLSG                   & 30.21      & 66.22     & 38.41       & 66.70   \\  \bottomrule       
\end{tabular}
\end{table}

\end{document}